\documentclass[%
 reprint,
 amsmath,amssymb,
 aps,
prb,
onecolumn]{revtex4-2}

\usepackage{graphicx}
\usepackage{dcolumn}
\usepackage{bm}
\usepackage{color}
\usepackage{float}
\usepackage{subcaption}
\usepackage{ragged2e}
\usepackage[unicode=true,pdfusetitle,
 bookmarks=true,bookmarksnumbered=false,bookmarksopen=false,
 breaklinks=false,pdfborder={0 0 1},backref=false,colorlinks=false]
 {hyperref}
 \hypersetup{colorlinks,allcolors=blue,breaklinks=true}
\newcommand{\A}[1]{\textcolor{red}{Amir: #1}}

\usepackage{pdfpages}
\makeatletter
\AtBeginDocument{\let\LS@rot\@undefined}
\makeatother

\begin{document}



# Reentrant superconductivity and superconductor-to-insulator transition in a naturally occurring Josephson junction array tuned by RF power

**S. Avraham, S. Sankar, S. Sandik, A. Burshtein, M. Goldstein, E. Sela, and Y. Dagan**[*]
*School of Physics and Astronomy, Tel Aviv University, Tel Aviv 6997801, Israel*

Superconductivity, characterized by dissipationless current flow with flux expulsion or quantization, is usually muted when the magnetic field or the temperature is sufficiently high. However, in rare instances, superconductivity can reappear upon increasing the temperature or magnetic field, a phenomenon known as reentrant superconductivity. It usually emerges from competing orders in strongly correlated materials. Here we demonstrate reentrant superconductivity as a function of both temperature and magnetic field, tuned by radio frequency (RF) power in a relatively simple system: granular aluminum (grAl), which exhibits the properties of a naturally occurring Josephson junction array. At low temperatures, giant Shapiro steps emerge, exhibiting characteristics of a single Josephson junction. Coherent phase locking across the array's multiple junctions amplifies the quantized voltage, enabling tunability at radio frequencies, as observed in artificially designed Josephson arrays. We show that our system can be tuned from a coherent superconducting (stiff-phase) to an insulating (phase-fluctuating) state using RF power. We propose that the RF power modulates the Josephson coupling energy, $E_J$. Remarkably, at elevated temperatures, the screening of the electron charge suppresses the charging energy, causing superconductivity to reappear. This many-body effect cannot be described within a single junction framework and involves many-body correlations. Our system can therefore be tuned to observe both the single-junction regime and many-body correlation effects, serving as a quantum simulator for complex phenomena in condensed matter physics.

---

[*] Corresponding author: yodagan@tauex.tau.ac.il

Complex behavior can arise in large networks governed by simple rules[1], a principle observed across physics in phenomena such as spontaneous magnetization in Ising models, phase transitions in percolation theory, and synchronization in coupled oscillators[2]. One prominent example of a quantum network exhibiting such emergent behavior is the Josephson junction (JJ) array, which offers a rich platform for exploring collective quantum dynamics[3]. In this correlated quantum system, the relevant energy scales are the Josephson coupling across the junction $E_J = \frac{\hbar I_c}{2e}$, which determines the confinement of the Josephson phase, and the charging energy $E_C = \frac{e^2}{2C}$, which determines the strength of Coulomb interactions, with $I_c$ being the Junction's critical current and $C$ the capacitance. The interplay between these energy scales yields a superconductor-to-insulator quantum phase transition[4], obtained for $E_J \simeq E_C$.

Reentrant superconductivity is a phenomenon in which the superconducting phase unexpectedly reappears from a dissipative state when increasing a tuning parameter of the system, such as temperature, magnetic field, and pressure. It has been observed in systems including superconductor-ferromagnet compounds[5–8] and layered structures[9,10], bismuthates [11], heavy-fermions[12,13], Kagome materials[14], topological insulator Josephson junctions[15], and twisted moiré materials[16]. This phenomenon often arises from competing orders, such as ferromagnetism and superconductivity in ErRh$_4$B$_4$ and HoMo$_6$S$_8$ [5,8], charge density waves and superconductivity in ZrTe$_2$ [17], or different pairing mechanisms in twisted-trilayer graphene[18]. Reentrant superconductivity has also been explored through gauge-gravity duality, where black hole stability serves as a probe of holographic superconductivity in the dual non-relativistic field theory[19].

Theoretically, it has been shown that within a specific regime of the quantum coupling ratio, $E_J/E_C$, reentrant superconductivity can emerge in JJ arrays[20]. At moderate temperatures, quasiparticle screening reduces $E_C$, enabling charge fluctuations and the reappearance of superconductivity. As the temperature decreases, this screening effect diminishes, restoring $E_C$ and driving the system back into a phase-fluctuating, dissipative state[20,21]. Despite its theoretical foundation, this reentrant behavior has remained experimentally elusive due to limitations in fabricating artificial JJ arrays with the required precision.

Various methods have been employed to tune $E_J/E_C$ in JJ arrays. For instance, varying the insulating barrier thickness to tune junction resistance[22], tuning junction critical current with a magnetic field[23], gating a two-dimensional electron gas placed in a proximity to the array to affect its screening[24], and electron depletion in semiconductor-superconductor hybrid devices[25]. While these methods could effectively tune $E_J/E_C$ across the superconductor-to-insulator transition, only traces for quasi-reentrant behavior were observed experimentally[22,26]. A clear reappearance of a dissipationless superconducting state in a JJ array is still lacking.

We have employed a new approach for realizing a JJ array and for tuning its parameters. We utilize a naturally occurring JJ array made of granular aluminum (grAl). This material consists of superconducting islands coupled through insulating interlayers[27] (See Figure 1c for illustration). The grain size and the barrier strength are controlled by the deposition conditions[28,29]. The system features nanoscale grain size, making the strong quantum fluctuations regimes of both charge and phase accessible[30]. Compared to artificial JJ arrays, the significantly smaller grain size results in a longer screening length, enhancing correlations between junctions across the array. Additionally, the electronic level spacing within each island

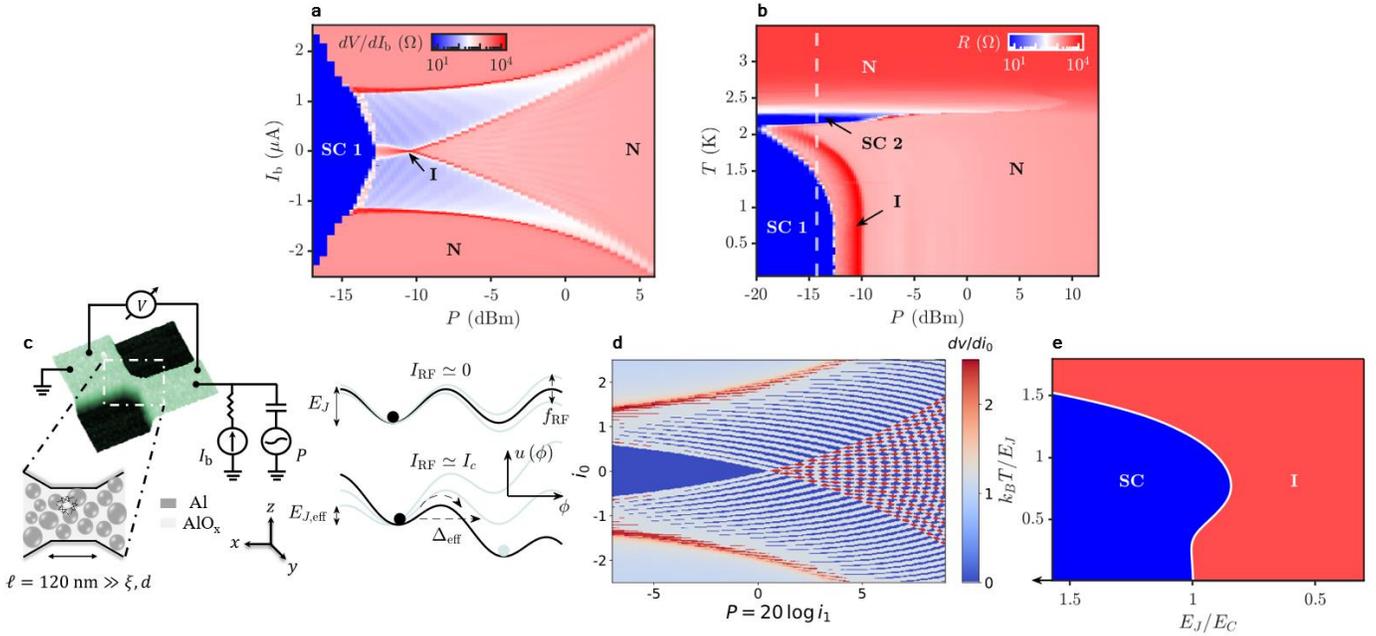

**Fig. 1 | Summary of main results. a,** Differential resistance $dV/dI_b$ as function of bias current $I_b$ and RF power $P$, demonstrating RF tuning of the system. For low $I_b$ the system is tuned through the superconductor (SC 1), the insulator (I), and the normal (N) regimes at constant frequency and temperature $f_{RF} = 24.5$ MHz, $T = 50$ mK. At higher currents, giant Shapiro steps are observed. **b,** The phase diagram, showing the zero-bias resistance $R$ as a function of temperature and RF power, obtained at $I_b = 0$. Reentrant superconductivity (SC 2) is observed at high temperatures and finite RF power. The superconductor (SC 1) to insulator (I) to normal (N) transition as a function of $P$ at constant $T$ is also observed. **c,** Scheme of the measurement setup is illustrated on an atomic force microscope (AFM) image, $I_b$ is the DC bias current and $P$ is the RF power, while the DC voltage $V$ is measured. Enlarged is a schematic illustration of the grAl nanobridge, with $\ell$, $d$, and $\xi$ the bridge length, mean grain size, and the Ginzburg-Landau coherence length, respectively. Right insert-Illustration of quantum phase slips induced by RF current in the inter-grain Josephson junction. For small RF currents, the phase is confined to a potential well, while the system is superconducting. By shaking the potential with considerable RF current $I_{RF}$ at frequency $f_{RF}$, the potential barrier of height $E_{J,eff}$ is reduced, inducing tunneling with enhanced rate $\Delta_{eff}$, while the system is insulating. **d,** Calculation of $dv/di_0$ as a function of $i_0$ and $P$ for a single Josephson junction using $\Omega = 0.1$ and $\beta = 1$. **e,** Phase diagram of a Josephson junction array with electrostatic screening length $\lambda_e/a = 10$ and charging energy $E_C$, predicting reentrant superconductivity at high temperatures (adapted from ref. [26]).

exceeds both $E_C$ and the superconducting gap[31–33], making the system less sensitive to microscopic details[34]. Therefore, it is a promising building block for high kinetic inductance and low-loss quantum devices[34]. Finally, to control the Josephson parameters, we adapted a new approach by application of radio frequency (RF) power to tune the critical current $I_c$ [35], thereby varying $E_J$. The drive frequency should be much smaller compared to the characteristic frequency of the junction $\omega_p \sim \sqrt{E_J E_C}$, so it is mainly $I_c$ that depends on the RF amplitude[36] (See Fig. 1c for illustration of the set-up and our interpretation for the effect of the RF power

on junction properties and the quantum phase slip rate). Utilizing the unique properties of grAl and our new tuning approach, we can tune through the various regimes of the device. We realize a superconductor-to-insulator transition as a function of RF power with a strong insulating state. This insulator has ten times higher resistance compared with the device normal resistance and the derivative of the device resistance with respect to temperature is negative (see Supplementary Information Fig. S4). In addition, in the insulator regime we observe a pronounced temperature dependent reentrant superconductivity. We explain our data using two regimes: at low temperatures, the system is described as a phase locked JJ array featuring giant Shapiro steps, obeying single-junction-type physics. At high temperatures, the reentrant behavior is described within the many-body correlated picture suggested by Efetov[20], where reentrance is induced by the screening of long-range interactions. Our findings emphasize the potential of using granular superconductors as controlled platforms for developing novel quantum devices and exploring emergent quantum phenomena in strongly correlated systems.

We study a JJ array implemented using a superconducting nanobridge fabricated from a grAl thin film (Fig. 1c). The nanobridge has a length of $\ell \simeq 120$ nm, width $w \simeq 200$ nm, and thickness $t \simeq 55$ nm. These dimensions are significantly larger than the grain size of grAl, $d = 2 \pm 0.5$ nm,[37] the inter-grain spacing ($s \simeq 1$ nm), and the superconducting coherence length ($\xi \simeq 10$ nm) [38]. Additionally, the bridge dimensions are much smaller than the in-plane magnetic penetration depth ($\lambda_m \simeq 1$ μm) [39], resulting in a relatively uniform current distribution along the array[40]. Naively, the bridge would consist of approximately $N_x \times N_y \times N_z \simeq 40 \times 67 \times 18$ inter-grain JJs. However, the effective number of junctions can differ significantly due to additional contributions coming from the contact region[41].

We begin by demonstrating the tunability of the system at a low temperature of $T = 50$ mK as shown in Fig. 1a. The differential resistance $dV/dI_b$ is shown to be tunable by the RF power $P$ and the bias current $I_b$. Below a critical power $P_c$ ($P < P_c = -13$ dBm), the superconducting state (blue region) is observed for low bias currents. Upon increasing the power at zero bias ($I_b \simeq 0$), a peak in the differential resistance emerges. This peak reaches its maximum height and minimum width at $P_I = -10.5$ dBm, which we define as the insulating state (I). We note that the zero-bias resistance $R_I$ is an order of magnitude larger than the resistance in the normal state $R_N$, obtained at the highest DC currents and RF powers where superconductivity is completely quenched. Furthermore, when raising the temperature, $R_I$ decreases, a typical behavior for an insulating state. Previous studies on JJ arrays [42,43] and single junctions[44] have reported power-induced transitions to a resistive state characterized by Shapiro step behavior. In contrast, the power-driven transition to an insulating state observed here is significantly more pronounced.

Upon increasing $I_b$ and $P$ further, clear oscillations of $dV/dI_b$ are observed. We interpret these oscillations as giant Shapiro steps, as evidenced from our further analysis in Fig. 3. These Shapiro steps demonstrate that the low temperature dependence of the system can be described within an effective single JJ model as described in the theoretical analysis below and depicted in Fig. 1d.

We now look at the $P - T$ phase diagram of the system, shown in Fig. 1b. We observe reentrant superconductivity as a function of temperature; *this is our main result*. This phenomenon has been predicted by Efetov for an array of JJ's as depicted in Fig. 1e. In our case we conjecture

that a finite RF power is needed for carefully tuning $E_J$ for one to be able to observe the reentrance behavior as will be discussed further below.

For illustration, let us first follow the dashed line in Fig. 1b at $P = -14.25$ dBm: at low temperatures, the system is superconducting with $R = 0$. As the temperature increases above 1.5 K, $R$ goes through a peak: this is the insulating state. Upon increasing the temperature further, superconductivity surprisingly reappears, this being the reentrant superconducting phase SC 2. Above 2.5 K, the normal state N of the system is observed. As noted above, the normal state resistance of the array is an order of magnitude smaller than the insulating state resistance.

The various regimes of the system, SC 1, I, SC 2, and N, are indicated on the phase diagram. Remarkably, we observe reentrant superconductivity as a function of temperature. For $P < P_c$, the system undergoes from low $T$ superconductivity (SC 1) to high $T$ superconductivity (SC 2), while crossing through the insulating (I) regime. In the insulating regime, $R(T)$ is first rising with lowering $T$, then saturating at low $T$. This saturation at the lowest temperatures may not be an intrinsic property of the system but a result of the electronic heat generated by the RF power [45] (see Supplementary Information Sec. V.B).

To investigate the effect of RF power and bias current, we fix the RF power at $P = -14.25$ dBm, and plot the differential resistance $dV/dI_b$ as a function of bias current $I_b$ and temperature $T$ (see Fig. 2a-b). As the temperature increases, the critical current $I_c$ of the superconducting state SC 1 decreases rapidly and vanishes completely at $T = 1.35$ K. Upon increasing the temperature further, the resistance rises, reaching a peak in the zero-bias resistance around $T = 1.8$ K.

Along the $I_b$ axis, (i.e., with increasing bias current), we observe oscillations in $dV/dI_b$ (Shapiro steps). At $T = 2.16$ K, the critical current $I_c$ reappears, signaling the onset of a reentrant superconducting phase (SC 2), which forms a dome-shaped region in the $I_b - T$ diagram at fixed RF power.

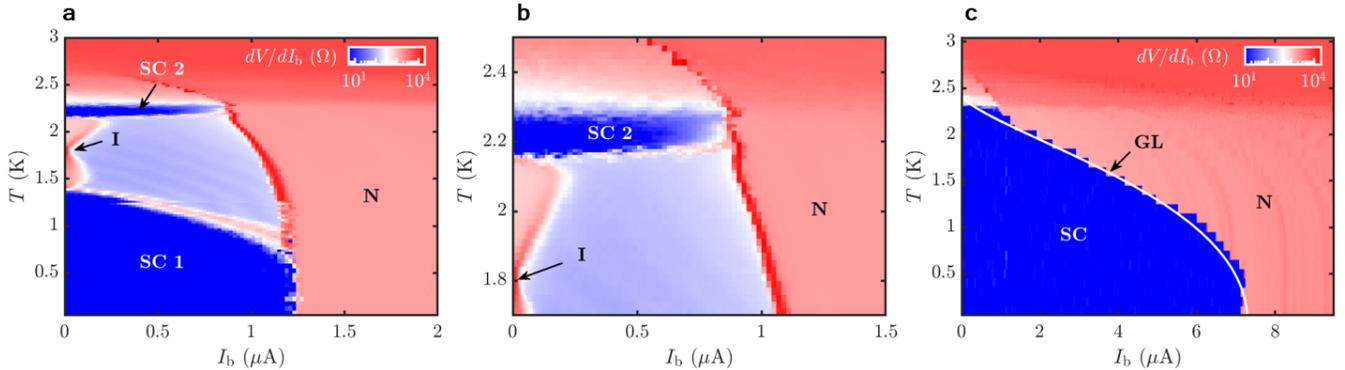

**Fig. 2 | Temperature dependence of the $V(I_b)$ characteristics.** $dV/dI_b$ as a function of $I_b$ and $T$. **a,** Measurement at fixed RF power and frequency, $P = -14.25$ dBm, $f_{RF} = 24.5$ MHz, demonstrating the various states of the system across the phase diagram. **b,** Enlarged view of **a** showing the insulating state and the reentrant superconducting dome appearing at high $T$. **c,** Measurement in the absence of RF power, showing conventional $I_b$ and $T$ dependencies. The solid curve represents a fit according to Ginzburg-Landau theory.

Fig. 2b, highlights this reentrant regime, where SC 2 is clearly visible. For comparison, Fig. 2c shows measurements taken without RF power, displaying a conventional $I_c(T)$ dependence without reentrance. As discussed further below, we propose that the presence of RF power is essential for fine-tuning the effective coupling energy $E_J$.

The dependence of the system response on the RF frequency, $f_{RF}$, is explored in Fig. 3. We study the differential resistance $dV/dI_b$ as a function of $I_b$ and $f_{RF}$, focusing on the insulating state, Fig. 3a-b. $dV/dI_b$ exhibits damped oscillations as a function of $1/f_{RF}$, Fig. 3a. The oscillations are smeared at low frequencies, while the insulator zero-bias peak is enhanced (see inset), suggesting it has a different origin. Looking at the $dV/dI_b$ in the $f_{RF} - V$ plane, Fig. 3b, we reveal the origin of the oscillations. They follow the giant Shapiro steps relation[42], $V_n = n N_j \phi_0 f_{RF}$. Here $n$ is an integer, $\phi_0$ is the flux quantum, and $N_j$ is the number of phase locked junctions connected in series. Using the frequency dependence, we obtain contributions from $N_j = 430$ junctions, which exceeds the number of junctions in the nanobridge ($N_x$). This difference can be attributed to an extended phase drop across the device[41], which comprises the nanobridge in series with long grAl contacts.

At a given frequency, the first consecutive Shapiro steps are missing. As the frequency decreases, more steps are hidden by the insulator peak, and only higher harmonic steps are visible. Fig. 3c shows the giant Shapiro steps observed at $f_{RF} = 67.5$ MHz. They clearly illustrate the even-odd effect expected for the current step width[46]. They are characterized by a finite voltage width expected for arrays with minuscule disorder[47] (see Supplementary Information Sec. V.C).

We extend our investigation by examining the RF response of the system under an external magnetic field (Fig. 4). At low temperatures and at zero fields, we define $P_c = -13$ dBm and $P_I = -10.5$ dBm at the onset of SC 1 and the insulating (I) regimes, respectively. They exhibit

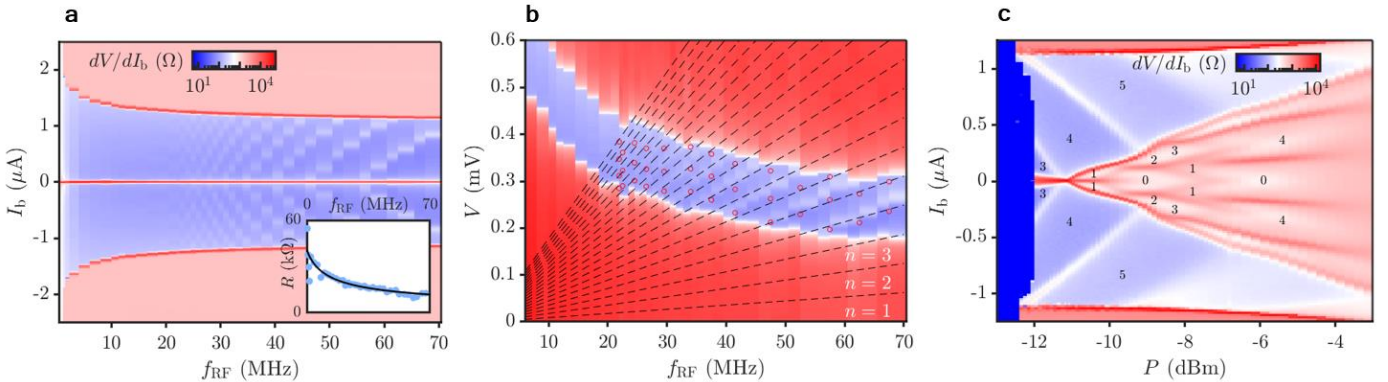

**Fig. 3 | Frequency dependence. a**, $dV/dI_b$ as a function of $I_b$ and $f_{RF}$, taken at power levels corresponding to the I state (for more details, see Supplementary Information Sec. V.B). The inset shows the zero-bias resistance $R$ vs. $f_{RF}$. The solid line used as a guide to the eye and represents $R \sim (f_{RF} + f_0)^{-1/2}$ behavior. **b**, $dV/dI_b$ in the $V - f_{RF}$ plane where minima points are indicated by red circles. The voltage is evaluated by numerical integration and linear interpolation of the $dV/dI_b$ data in **a**. The dashed lines follow the giant Shapiro steps relation $V_n = n N_j \phi_0 f_{RF}$ for $N_j = 430$. **c**, $dV/dI_b$ as a function of $I_b$ and $P$ at $f_{RF} = 67.5$ MHz. The step number $n$ is indicated.

non-monotonic modulation by the out-of-plane magnetic field $\mu_0 H_\perp$ (Fig. 4a). The oscillations of the critical power $P_c$ are reminiscent of a Fraunhofer-type interference. The oscillatory, non-monotonic behavior of $P_I$ as a function of magnetic field, along with the strong suppression of $R_I$ at small $\mu_0 H_\perp$, suggests that the insulating phase emerges from the superconducting state.

In contrast, when an in-plane magnetic field $\mu_0 H_\parallel$, aligned with the current direction, is applied, both $P_c$ and $P_I$ decrease monotonically, while $R_I$ is suppressed only at significantly higher fields (Fig. 4b). In both magnetic field orientations, and within the power range $P_c < P < P_I$, distinct resistance oscillations are observed. These oscillations correspond to the missing Shapiro steps, which reemerge once the zero-bias peak associated with the insulating state is suppressed by the magnetic field.

Remarkably, at elevated temperatures, we observe signatures of field-induced reentrant superconductivity, most prominently under in-plane field conditions (Fig. 4c-d). This reentrance may be consistent with Efetov's picture: here, quasiparticles would be produced by pair-breaking[48,49], enhancing screening, when the applied magnetic field is close to the upper critical field[38].

## Theoretical model

For simplicity, we conceptualize our system as a stack of many parallel one-dimensional chains of JJs. This model captures the Shapiro steps observed in Fig. 1a at low temperature and RF frequencies well below $\omega_p$, using effective single-junction physics, as we now explain. The semiclassical equation of motion for each junction is given by the model of a resistively and capacitively shunted junction (RCSJ) under simultaneous DC and RF bias,

$$\beta \frac{\partial^2 \phi}{\partial \tau^2} + \frac{\partial \phi}{\partial \tau} + \sin \phi = i_0 + i_1 \sin(\Omega \tau), \tag{1}$$

where $\beta = 2\pi R^2 C I_c / \phi_0$ is set by the normal resistance $R$, the capacitance $C$, and the critical current $I_c$ of the junction. The time $t$ has been rescaled by $t_R = \phi_0 / (2\pi R I_c)$ to define the dimensionless parameter $\tau = t/t_R$, $\Omega = \omega_{RF} t_R$ is the rescaled RF frequency, and $i_0 = I_b/I_c$, $i_1 = I_{RF}/I_c$ are the rescaled DC and RF currents, respectively (the latter being proportional to the square root of the RF power, $I_{RF} \sim \sqrt{P}$). The voltage drop across the junction is given by $V = \phi_0 \partial_t \phi / 2\pi$. We assume that the DC and RF currents flowing through the sample are distributed equally between the one-dimensional chains; then, within each chain, the RCSJ equations decouple, and the total voltage drop across the chain is given by the voltage drop across a single junction multiplied by the number of junctions in each chain.

In Fig. 1d, we present calculations of $dv/di_0$ as a function of the rescaled DC current $i_0$ and RF power $P$, where $v = \partial_\tau \phi = 2\pi t_R V / \phi_0$ is the rescaled voltage, obtained for $\Omega = 0.1$ and $\beta = 1$ (see Supplementary Information Sec. III for extended discussion and other parameter regimes). While the model is simplified, it captures the Shapiro steps observed in Fig. 1a; this is in line with previous studies of JJ arrays[42]. Moreover, it accounts for the oscillatory behavior seen as a function of temperature in Fig. 2a: temperature reduces the critical current such that $i_{0,1} = I_{b,RF}/I_c(T)$ in Eq. (1) become temperature-dependent, leading to the emergence of Shapiro steps.

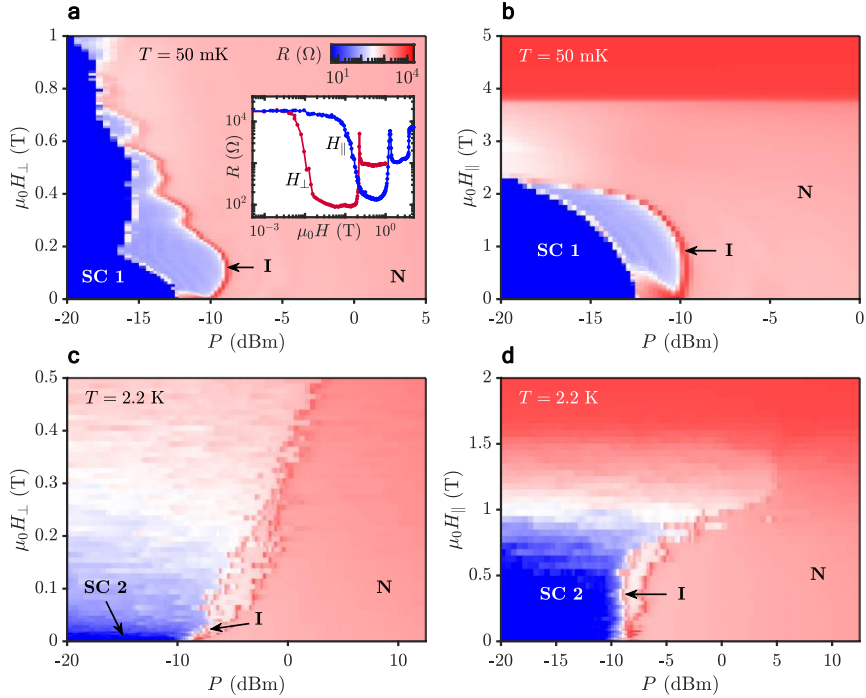

**Fig. 4 | Effects of magnetic field.** Zero bias resistance as a function of magnetic field and RF power at $f_{RF} = 24.5\ MHz$. **a-b,** Measurements at $T = 50\ mK$, in the out-of-plane field $\mu_0 H_\perp$ and in-plane field $\mu_0 H_\parallel$ configurations, respectively. The inset shows line cuts at $P = -10.5\ dBm$, demonstrating strong suppression of the insulator resistance at small out-of-plane magnetic fields. **c-d,** Measurements at $T = 2.2\ K$ showing reentrance of the high temperature superconducting phase SC 2 as a function of magnetic field.

While the RCSJ model accounts for the giant Shapiro steps in Fig. 1a, it does not capture the observed zero-bias peak $R_I$. As mentioned above, it is an order of magnitude larger than $R_N$ and the oscillations amplitude, which should be of the same order of magnitude as the zero-bias peak according to the RCSJ model. The qualitative behavior of the zero-bias peak and the oscillations differ as well: $R_I$ increases at small RF frequencies, while the oscillations are suppressed (Fig. 3a). Moreover, $R_I$ quickly diminishes at small out-of-plane magnetic fields, whereas the oscillations are only shifted (Fig. 4a). These observations suggest that $R_I$ has a different origin. To this end, we note that insulating behavior in JJs and JJ arrays[4,50] is often discussed in the context of the Bloch band picture[51], where phase slips proliferate and the almost-classical variable is the quasicharge; this picture is clearly incompatible with the RCSJ model, where the phase is stiff. We discuss this possibility in the Supplementary Information and offer a possible explanation for the qualitative dependence of the zero-bias peak on the RF power and frequency using the Bloch band picture, revealing a non-trivial dependence of the phase slip rate on the RF current (see Supplementary Information Sec. I). A full analysis stitching together these two regimes is beyond the scope of our calculations and marks an exciting prospect for future research.

The reentrant superconductivity observed at elevated temperatures is a many-body effect which is beyond the scope of the single-junction RCSJ equation. A suitable framework is provided by the Efetov model[20], that predicts the screening of long-range Coulomb interactions due to

Cooper-pair excitations at finite temperatures, reducing the effective charging energy $E_C$ of the array junctions. According to this picture, increasing the temperature does not necessarily lead to a decrease in the ratio $E_J/E_C$, which determines the onset of superconductivity in a two-dimensional array[26] – while $E_J \propto I_c$ decreases, the ratio $E_J/E_C$ may grow due to the decrease in $E_C$, resulting in reentrant superconductivity. We note that the reentrance could be washed out if the electrostatic screening length, $\lambda_e$, is too small. Where $\lambda_e = a\sqrt{C/C_0}$ is defined by the array mutual capacitance $C = e^2/(2E_C)$, the self-capacitance $C_0$ of each aluminum grain to the ground, and the array spacing $a$ [26]. However, in our experiment, we estimate $\lambda_e/a \sim 10^2$, where reentrance is expected to be clearly visible, as illustrated in Fig. 1e. Furthermore, our system was shown to be close to a Mott transition[52], hinting that strong Coulomb interactions are indeed present. Fig. 2a shows that reentrance occurs from both the insulating and normal states; we phenomenologically explore the latter in Supplementary Information Sec. IV by using the effective $E_C$, reduced by the many-body screening, in the RCSJ equation.

We reiterate that the reentrance appears only upon the application of RF current and does not occur using DC current. Indeed, a DC current tilts the washboard potential, and thermal fluctuations push the phase towards a preferred direction, giving rise to a finite resistance. On the contrary, RF current shakes the washboard potential periodically, such that the phase has no preferred direction on average. Within the RCSJ picture, where the phase is an almost-classical variable, the periodic shaking of the potential leads to an effective decrease in $E_J$, as illustrated in Fig. 1c and shown in Supplementary Information Sec. II. The RF power therefore controls the effective junction parameters, providing the necessary knob to explore the Efetov phase diagram in Fig. 1e and observe the reentrance of superconductivity. We note that the dependence of the junction parameters on the RF power could be significantly altered by the effect of quantum fluctuations – we elaborate on this point in the Supplementary Information.

In summary, we study a naturally occurring nano-scale JJ array tuned by RF power. We demonstrate that the system can be tuned from the giant Shapiro-step regime to an insulating state, where the resistance increases to ten times the normal-state resistance. The oscillatory dependence of the array resistance on the perpendicular magnetic field in the insulating regime suggests that this insulating state emerges from the superconducting phase. The most striking phenomenon observed is reentrant superconductivity, which appears at elevated temperatures or under high magnetic fields. We propose that the RF current modulates key junction parameters—such as the Josephson coupling and the phase slip rate—thereby enabling the observation of reentrant behavior. While the coherent state is well explained by the RCSJ model, the crossover to the insulating regime and the emergence of reentrant superconductivity presents an exciting opportunity for advancing the theory of JJ arrays.

## Methods

### Device fabrication

The naturally occurring JJ arrays have been implemented by granular aluminum nanobridge devices. The nanobridges were obtained using standard liftoff electron beam lithography. Device dimensions were examined by atomic force microscopy. Granular aluminum thin films were thermally evaporated on R-plane sapphire substrates held at liquid nitrogen temperature.

Compared to films deposited on substrates held at room temperature, this approach reduces the mean grain size and narrows the grain size distribution[29]. For the nanobridge device presented in the main text, the film resistivity is $\rho = 2800\ \mu\Omega$cm, estimated by measuring an on-chip Hall bar device at room temperature.

**Measurement setup**

Measurements were conducted in a dilution refrigerator with a base temperature of 20 mK and a (8 T, 1 T, 1 T) superconducting vector magnet. DC transport measurements were taken while simultaneously applying RF signals characterized by a power, $P$ (dBm), defined in the signal generator output. The measurement setup is illustrated in Supplementary Information Fig. S6. The DC lines are filtered using a multi-stage RF and DC low-pass filter held on top of the mixing chamber plate, followed by DC low-pass filtering on the sample board. The RF lines are attenuated at different stages of the dilution refrigerator and are coupled to the device through on-board RF bias-tees. The bias tees consist of a 50 k$\Omega$ DC bias resistor and a 22 nF RF capacitor. Differential resistance, $dV/dI_b$, data were taken by applying a DC current $I_b$ in addition to a small AC current $dI_b = 10$ nA at a frequency of 7.189 Hz, using a standard lock-in technique.

## Acknowledgments

This paper is dedicated to the memory of Guy Deutscher 1936-2024.

## Author Contributions

S. Avraham fabricated the device. S. Avraham and S. Sandik conducted the measurements. S. Sankar, A. Burshtein, M. Goldstein, and E. Sela developed the model and performed the simulations. S. Avraham and Y. Dagan analyzed the data. All authors discussed the data and participated in writing the paper.

## References

1.      Strogatz, S. H. Exploring complex networks. *Nature* **410**, 268–276 (2001).
2.      Fradkin, E. *Field Theories of Condensed Matter Physics, Second Edition*. *Field Theories of Condensed Matter Physics, Second Edition* vol. 9780521764445 (2010).
3.      Fazio, R. & Van Der Zant, H. *Quantum Phase Transitions and Vortex Dynamics in Superconducting Networks*. *Physics Reports* vol. 355 (2001).
4.      Takahide, Y., Yagi, R., Kanda, A., Ootuka, Y. & Kobayashi, S. I. Superconductor-Insulator Transition in a Two-Dimensional Array of Resistively Shunted Small Josephson Junctions. *Phys Rev Lett* **85**, 1974 (2000).

5.      Fertig, W. A. *et al.* Destruction of Superconductivity at the Onset of Long-Range Magnetic Order in the Compound ErRh4B4. *Phys Rev Lett* **38**, 987 (1977).

6.      Lynn, J. W., Gotaas, J. A., Shelton, R. N., Horng, H. E. & Glinka, C. J. Magnetic and superconducting properties of holmium-rich (Er1-x Hox)Rh4B4. *Phys Rev B* **31**, 5756 (1985).

7.      Miclea, C. F. *et al.* Evidence for a reentrant superconducting state in EuFe2As2 under pressure. *Phys Rev B Condens Matter Mater Phys* **79**, (2009).

8.      Lynn, J. W., Shirane, G., Thomlinson, W., Shelton, R. N. & Moncton, D. E. Magnetic properties of the reentrant ferromagnetic superconductor HoMo6S8. *Phys Rev B* **24**, 3817 (1981).

9.      Garifullin, I. A. *et al.* Re-entrant superconductivity in the superconductor/ferromagnet V/Fe layered system. *Phys Rev B Condens Matter Mater Phys* **66**, 205051–205054 (2002).

10.     Zdravkov, V. *et al.* Reentrant superconductivity in Nb/Cu1-xNix bilayers. *Phys Rev Lett* **97**, (2006).

11.     Lin, T. H. *et al.* Observation of a reentrant superconducting resistive transition in granular BaPb0.75Bi0.25O3 superconductor. *Phys Rev B* **29**, 1493 (1983).

12.     Knebel, G. *et al.* Field-reentrant superconductivity close to a metamagnetic transition in the heavy-fermion superconductor Ute2. *J Physical Soc Japan* **88**, 063707 (2019).

13.     Ran, S. *et al.* Extreme magnetic field-boosted superconductivity. *Nature Physics* vol. 15 1250–1254 Preprint at https://doi.org/10.1038/s41567-019-0670-x (2019).

14.     Chen, X. *et al.* Highly robust reentrant superconductivity in csv3sb5 under pressure. *Chinese Physics Letters* **38**, (2021).

15.     Mandal, P. *et al.* Magnetically tunable supercurrent in dilute magnetic topological insulator-based Josephson junctions. *Nat Phys* **20**, 984–990 (2024).

16.     Cao, Y., Park, J. M., Watanabe, K., Taniguchi, T. & Jarillo-Herrero, P. Pauli-limit violation and re-entrant superconductivity in moiré graphene. *Nature* **595**, 526–531 (2021).

17.     Yomo, R., Yamaya, K., Abliz, M., Hedo, M. & Uwatoko, Y. Pressure effect on competition between charge density wave and superconductivity in ZrTe3: Appearance of pressure-induced reentrant superconductivity. *Phys Rev B Condens Matter Mater Phys* **71**, 132508 (2005).

18.     Lake, E. & Senthil, T. Reentrant superconductivity through a quantum Lifshitz transition in twisted trilayer graphene. *Phys Rev B* **104**, (2021).

19.     Cremonesi, S., Melnikov, D. & Oz, Y. Stability of asymptotically Schrödinger RN black hole and superconductivity. *Journal of High Energy Physics* **2010**, 1–17 (2010).

20.     Efetov, K. B. Phase transition in granulated superconductors. *Zh. Eksp. Teor. Fiz.* **78**, 2017 (1980).

21.     Feigel'man, M. V, Korshunov, S. E. & Pugachev, A. B. Parity effect and charge-binding transition in submicron Josephson junction arrays. *Zh. Eksp. Teor. Fiz.* **65**, 541 (1997).

22.     Van Der Zant, H. S. J., Elion, W. J., Geerligs, L. J. & Mooij, J. E. Quantum phase transitions in two dimensions: Experiments in Josephson-junction arrays. *Phys Rev B* **54**, 10081 (1996).

23.     Kuo, W. & Chen, C. D. Scaling Analysis of Magnetic-Field-Tuned Phase Transitions in One-Dimensional Josephson Junction Arrays. *Phys Rev Lett* **87**, 186804 (2001).

24.     Rimberg, A. J. *et al.* Dissipation-Driven Superconductor-Insulator Transition in a Two-Dimensional Josephson-Junction Array. *Phys Rev Lett* **78**, 2632 (1997).

25. Bøttcher, C. G. L. *et al.* Superconducting, insulating and anomalous metallic regimes in a gated two-dimensional semiconductor–superconductor array. *Nat Phys* **14**, 1138–1144 (2018).

26. Newrock, R. S., Lobb, C. J., Geigenmüller, U. & Octavio, M. The Two-Dimensional Physics of Josephson Junction Arrays. in *Solid State Physics - Advances in Research and Applications* vol. 54 (2000).

27. Deutscher, G., Gershenson, M., Grünbaum, E. & Imry, Y. Granular Superconducting Films. *Journal of Vacuum Science and Technology* **10**, 697–701 (1973).

28. Moshe, A. G., Farber, E. & Deutscher, G. Optical conductivity of granular aluminum films near the Mott metal-to-insulator transition. *Phys Rev B* **99**, 224503 (2019).

29. Glezer Moshe, A., Tuvia, G., Avraham, S., Farber, E. & Deutscher, G. Tunneling study in granular aluminum near the Mott metal-to-insulator transition. *Phys Rev B* **104**, 054508 (2021).

30. Deutscher, G. Critical fluctuations in granular superconductors. *Phys Lett A* **35**, 28–29 (1971).

31. Deutscher, G. Granular Superconductivity: a Playground for Josephson, Anderson, Kondo, and Mott. *Journal of Superconductivity and Novel Magnetism* vol. 34 Preprint at https://doi.org/10.1007/s10948-020-05773-y (2021).

32. Bachar, N. *et al.* Kubo spins in nanoscale aluminum grains: A muon spin relaxation study. *Phys Rev B* **101**, 024424 (2020).

33. Pracht, U. S. *et al.* Optical signatures of the superconducting Goldstone mode in granular aluminum: Experiments and theory. *Phys Rev B* **96**, 094514 (2017).

34. Glezer Moshe, A., Farber, E. & Deutscher, G. Granular superconductors for high kinetic inductance and low loss quantum devices. *Appl Phys Lett* **117**, (2020).

35. Dayem, A. H. & Wiegand, J. J. Behavior of Thin-Film Superconducting Bridges in a Microwave Field. *Physical Review* **155**, 419 (1967).

36. Russer, P. Influence of microwave radiation on current-voltage characteristic of superconducting weak links. *J Appl Phys* **43**, 2008–2010 (1972).

37. Lerer, S., Bachar, N., Deutscher, G. & Dagan, Y. Nernst effect beyond the coherence critical field of a nanoscale granular superconductor. *Phys Rev B Condens Matter Mater Phys* **90**, (2014).

38. Glezer Moshe, A., Farber, E. & Deutscher, G. From orbital to Pauli-limited critical fields in granular aluminum films. *Phys Rev Res* **2**, (2020).

39. Abraham, D., Deutscher, G., Rosenbaum, R. & Wolf, S. Magnetic Penetration Depth in Granular Al-Al2O3 Films. *Journal de Physique Colloques* **39**, (1978).

40. Pearl, J. Current distribution in superconducting films carrying quantized fluxoids. *Appl Phys Lett* **5**, 65–66 (1964).

41. Vijay, R., Sau, J. D., Cohen, M. L. & Siddiqi, I. Optimizing anharmonicity in nanoscale weak link Josephson junction oscillators. *Phys Rev Lett* **103**, 087003 (2009).

42. Benz, S. P., Rzchowski, M. S., Tinkham, M. & Lobb, C. J. Fractional giant Shapiro steps and spatially correlated phase motion in 2D Josephson arrays. *Phys Rev Lett* **64**, 693 (1990).

43. Panghotra, R. *et al.* Giant fractional Shapiro steps in anisotropic Josephson junction arrays. *Commun Phys* **3**, 1–8 (2020).

44. Baars, P., Richter, A. & Merkt, U. Temperature and power dependence of Shapiro and Fiske step widths in Nb/InAs/Nb Josephson junctions. *Phys Rev B* **67**, 224501 (2003).

45. Tamir, I. *et al.* Sensitivity of the superconducting state in thin films. *Sci Adv* **5**, (2019).

46. Kvale, M., Hebboul, S. E. & Garland, J. C. Theory of even-odd step suppression in the RSJ model. *Physica B Condens Matter* **165–166**, 1585–1586 (1990).

47. Ravindran, K. *et al.* Frequency dependence of giant Shapiro steps in ordered and site-disordered proximity-coupled Josephson-junction arrays. *Phys Rev B* **53**, 5141 (1996).

48. Chandrasekhar, B. S. A NOTE ON THE MAXIMUM CRITICAL FIELD OF HIGH-FIELD SUPERCONDUCTORS. *Appl Phys Lett* **1**, 7–8 (1962).

49. Clogston, A. M. Upper limit for the critical field in hard superconductors. *Phys Rev Lett* **9**, (1962).

50. Penttilä, J. S., Parts, Ü., Hakonen, P. J., Paalanen, M. A. & Sonin, E. B. *'Superconductor-Insulator Transition' in a Single Josephson Junction.* (1999).

51. Likharev, K. K. & Zorin, A. B. *Theory of the Bloch-Wave Oscillations in Small Josephson Junctions. Journal of Low Temperature Physics* vol. 59.

52. Bachar, N. *et al.* Mott transition in granular aluminum. *Phys Rev B Condens Matter Mater Phys* **91**, (2015).



\title{Supplementary Information for: Reentrant superconductivity and superconductor-to-insulator transition in a naturally occurring Josephson junction array tuned by RF power}
\author{S. Avraham}
\author{S. Sankar}
\author{S. Sandik}
\author{A. Burshtein}
\author{M. Goldstein}
\author{E. Sela}
\author{Y. Dagan}
\email{Corresponding author: yodagan@tauex.tau.ac.il}
\affiliation{%
  School of Physics and Astronomy, Tel Aviv University, Tel Aviv 6997801, Israel}
\date{\today}
\maketitle
\section{The Bloch band picture}
As noted in the main text, the zero-bias peak $R_{\textrm{I}}$ in Fig.~1a of the main text cannot be captured within the RCSJ model. We speculate that its emergence may be attributed to the dual Bloch band picture. The periodic potential of a Josephson junction, $-E_J\cos\phi$, gives rise to Bloch wavefunctions that are $2\pi$-periodic in $\phi$ up to a phase that depends on the quasicharge $q$, which is analogous to the crystal momentum $k$ in solid state systems and is defined in the Brillouin zone $-e\le q\le e$. From the solid state analogy, we find energy bands $E_n(q)$ that depend periodically on $q$ and are given by the Mathieu characteristic values \cite{koch2007charge}, where $n=0,1,\ldots$ is the band index. In the experimentally-relevant limit $E_J \gg E_C$, the junction potential is in the tight-binding limit, and the lowest Bloch band is approximately given by $E_0(q)\approx-\Delta\cos(\pi q/e)$, where $\Delta \approx \frac{32}{2^{1/4}\sqrt{\pi}}(E_J^3E_C)^{1/4}e^{-\sqrt{8 E_J / E_C}}$ is the phase slips rate, which is the tunneling amplitude between adjacent minima in the cosine potential.

The superconducting phase $\phi$ and the quasicharge $q$ are conjugate variables and satisfy an uncertainty relation. If the quantum fluctuations in the quasicharge are small (corresponding to a fluctuating phase), the semiclassical equation of motion for a junction biased by DC and RF currents $I_{\textrm{b},\textrm{RF}}$ reads \cite{likharev1985bloch}
\begin{equation} \label{eq:q_eom}
    \dot q = -\frac{1}{R_{\textrm{qp}}}\frac{d E_0(q)}{dq} + I_{\textrm{b}} + I_{\textrm{RF}}\sin(\omega_{\textrm{RF}} t),
\end{equation}
where $R_{\textrm{qp}}$ is the quasiparticle resistance of the junction and $\omega_{\textrm{RF}}=2\pi f_{\textrm{RF}}$. Here we assume that the bias currents and the temperature are small enough such that the quasicharge is restricted to the lowest Bloch band, $n=0$. This equation is dual to the RCSJ equation (Eq.~(1) of the main text), with the absence of a second-derivative intertia term $\ddot{q}$; this term may be added with an inductor $L$ connected in series with the junction (dual to the parallel capacitance in the RCSJ model), but is unimportant in the following discussion and we omit it henceforth.

The analysis of Eq.~\eqref{eq:q_eom} is due to Likharev and Zorin \cite{likharev1985bloch}. Consider the case $I_{\textrm{RF}}=0$, and set $E_0(q) \approx -\Delta\cos(\pi q/e)$. At $I_{\textrm{b}} < I_{\textrm{th}}$ with $I_{\textrm{th}}=\pi\Delta/(eR_{\textrm{qp}})$, the solution for the quasicharge is stationary, $\dot q = 0$, and the current flows through the quasiparticle channel, such that the voltage across the junction is $V = I_{\textrm{b}} R_{\textrm{qp}}$. As $I_{\textrm{b}}$ exceeds the threshold $I_{\textrm{th}}$, the quasicharge is no longer stationary and enters the Bloch oscillations regime, less current flows through the quasiparticle channel, and the differential resistance $dV/dI_{\textrm{b}}$ is negative. One therefore expects a resistance peak around $I_{\textrm{b}}=0$ with a width $I_{\textrm{th}}$. Such a peak was observed for resistively-shunted junctions \cite{penttila1999superconductor}, and also for a two-dimensional Josephson array \cite{takahide2000superconductor}.

Let us now consider the effect of a small RF current on the resistance peak. Within linear response in $I_{\textrm{RF}}$, we write $q(t) = \tilde{q}(t) + \Re\{\delta q e^{i\omega_{\textrm{RF}} t}\}$, where $\tilde q(t)$ is the solution to Eq.~\eqref{eq:q_eom} in the absence of RF current. Plugging $q(t)$ to Eq.~\eqref{eq:q_eom} and expanding in $\delta q / \tilde q \ll 1$, we find
\begin{equation}
    \delta q = \frac{I_{\textrm{RF}}}{\omega_{\textrm{RF}} + i\frac{\pi^2 \Delta}{e^2 R_{\textrm{qp}}}\cos(\pi \tilde q/e)}.
\end{equation}
The correction $\delta q$ renormalizes the Bloch bandwidth $\Delta$. To see this, plug $q = \tilde q + \Re\{\delta q e^{i\omega_{\textrm{RF}} t}\}$ into $\Delta \sin(\pi q(t)/e)$:
\begin{equation}
    \Delta \sin(\pi q(t)/e) = \Delta J_0(\pi |\delta q|/e)\times \sin(\pi \tilde q/e) + \sum_{n\neq 0} J_n(\pi |\delta q|/e) e^{i n (\omega_{\textrm{RF}} t - \varphi)},
\end{equation}
where $J_n(x)$ is the $n$th Bessel function of the first kind, and $\varphi$ is the phase of $\delta q$. Integrating over a period of the RF current, the high-order Bessel functions average out, and we are left with a renormalized DC term:
\begin{equation}\label{eq:Delta_eff}
    \Delta_{\textrm{eff}} = \Delta J_0(\pi |\delta q|/e).
\end{equation}
Using $J_0(x \ll 1) \approx 1 - x^2/4$, we find that $\Delta_{\textrm{eff}}$ is reduced as the RF current is increased. This effect is expected to diminish at large frequencies, $\omega_{\textrm{RF}} \gg \pi^2 \Delta / (e^2 R_{\textrm{qp}})$. We note that the same conclusions may be derived by diagonalizing the Floquet Hamiltonian of a periodically-tilted cosine potential and extracting the width of the lowest Bloch band from the Floquet quasienergies \cite{gomez2013floquet}. Recalling that the Bloch bandwidth $\Delta$ is the tunneling rate in the cosine potential, this effect may be interpreted as the coherent destruction of tunneling \cite{grossmann1991coherent}.

In the experiment, the zero-bias resistance peak narrows as the RF power is increased, as shown in Fig.~1a of the main text. This is in line with the prediction of the Bloch band picture: decreasing $\Delta_{\textrm{eff}}$ corresponds to decreasing $I_{\textrm{th}} = \pi\Delta_{\textrm{eff}}/(eR_{\textrm{qp}})$. However, the rest of Fig.~1a is captured quite well by the RCSJ model, which cannot coexist with the Bloch band picture, since the two models correspond to two different regimes of quantum fluctuations in $\phi$ and $q$. This possibly suggests that a more complete analysis is required to explain the experimental observations, not only the low-temperature phase diagram, but also the reentrant superconductivity at finite temperatures.

\section{Controlling $E_J$ with RF current}
In this section, we demonstrate how the application of RF current leads to a reduction in $E_J$ within the RCSJ picture. Consider the semiclassical equation of motion for the phase,
\begin{equation}\label{eq:phi_eom}
    \frac{\phi_0}{2\pi}C\ddot{\phi} + \frac{\phi_0}{2\pi R}\dot\phi + I_c\sin(\phi) = I_{\textrm{b}} + I_{\textrm{RF}}\sin(\omega_{\textrm{RF}} t).
\end{equation}
Note that now $R$ is the normal state resistance of the junction and not the quasiparticle resistance in the Bloch band picture. Neglecting the capacitive intertia term, which is unimportant in this context, the equation for the phase $\phi$ is dual to Eq.~\eqref{eq:q_eom} for the charge, with the critical current $I_c$ replacing the phase slips rate $\Delta$. One may repeat the calculation from the previous section in the present situation: applying linear response in the RF current leads to a renormalization of the critical current,
\begin{equation}\label{eq:Ic_eff}
    \delta I_{c,\textrm{eff}} = I_c J_0(|\delta\phi|),
\end{equation}
with
\begin{equation}
    \delta\phi = \frac{R I_{\textrm{RF}}}{\frac{\phi_0}{2\pi}\omega_{\textrm{RF}}+i RI_c\cos(\tilde\phi)},
\end{equation}
where $\tilde\phi$ is the solution to Eq.~\eqref{eq:phi_eom} at $I_{\textrm{RF}}=0$. Thus, we find that $E_{J,\textrm{eff}} = \Phi_0 I_{c,\textrm{eff}}/(2\pi)$ decreases as $I_{\textrm{RF}}$ increases.

Note that Eqs.~\eqref{eq:Delta_eff} and \eqref{eq:Ic_eff} result from the Bloch band and RCSJ pictures, respectively, which correspond to different regimes of quantum fluctuations and hint at a non-trivial dependence of the junction parameters on the RF current. On the one hand, the Bloch band picture predicts that the phase slips rate is decreased as the RF current is increased. On the other hand, the RCSJ model predicts a decrease in $I_c$ and hence $E_J$ as well, corresponding to a larger phase slips rate $\Delta \sim e^{-\sqrt{8 E_J/E_C}}$. We conclude that the dependence of the junction parameters on the RF current is drastically altered by the regime of quantum fluctuations.

\section{Comparing the experiment with the RCSJ model}
In a simplified picture, the granular aluminum (grAl) sample can be thought of as a stack of many parallel  1D JJ chains. When current biased, the currents flowing through the parallel chains would be roughly the same since the long 1D chains are statistically similar. The individual JJs in the 1D chain are modeled as RCSJ junctions and so the 1D chain consists of RCSJ junctions in series. In the current bias case, the dynamics of the individual RCSJ junctions decouple with the individual dynamics described by Eq.~(1) in the main text. The total voltage across the chain is simply the sum of voltages across the individual junctions. The RCSJ parameters $\Omega=\phi_0 f_{\textrm{RF}}/(I_c R),\,\beta=2\pi R^2 C I_c/\phi_0$ of the individual junctions are expected to be very similar with some mean value and a small variance; the small variance is expected to result in smoothening of sharp features such as the Shapiro steps~\cite{Ravindran1996}. We attempt to estimate the RCSJ parameters (mean value) for the experimental system.

We can estimate $\Omega$ directly from Fig.~1a of the main text, where the RF drive is at frequency $f_{\textrm{RF}}= 24.5\ \textrm{MHz}$. The number of steps, $N_s$, that would occur for a given range of the normalized DC current, $\delta i$, is determined by $\Omega$: $N_s\sim \delta i /\Omega$. In the figure we can see around 3 steps between $I_{\textrm{b}}=0$ and $I_{\textrm{b}}=1\ \mathrm{\mu}$A. Then, using the zero-temperature critical current value for the sample (see Fig. 2c in the main text), $I_{c,\rm{tot}}$ $\sim 7\ \mathrm{\mu A}$, we get $\Omega\sim 0.05$. Now, let us calculate $\Omega$ using its defining relation, $\Omega=\phi_0 f_{\textrm{RF}}/(I_c R)$. Let $N_c$ denote the number of parallel chains and $N$ denote the number of RCSJ junctions in a chain. At large current bias, all the RCSJ junctions become purely ohmic, and then the total voltage across the sample is simply given by, $V=N R I_{\textrm{b}}/N_c$. From the frequency dependence of the giant Shapiro steps, we find $N=430$ in the main text. The critical current of an RCSJ junction is related to the total critical current of the sample by, $I_{c}=I_{c,\rm{tot}}/N_c$. Then we get,
\begin{equation}
\label{eq:icr}
    I_c R = \frac{I_{c,\rm{tot}}}{N}\frac{V}{I_\textrm{b}}, \quad \rm{for\, large\,}\,\textit{I}_{\textrm{b}}.
\end{equation}
The sample resistance at large bias is $V/I_{\textrm{b}} \sim 1\,\textrm{k}\Omega$. Then, plugging in the value of $I_{c,\rm{tot}}$ and $N$, we get, $I_c R \sim 0.01~\textrm{mV}$ and consequently $\Omega \sim 0.005$. This estimate is one order of magnitude smaller than the directly estimated $\Omega$ from Fig.~1a. The origin of this discrepancy is not fully clear. One possible cause could be that the superconducting coherence length $\sim 10~\textrm{nm}$ is greater than the grain size, and this could result in multiple grains combining to form an effective RCSJ junction. In that case, the factor $N=430$ reported in the main text would correspond to the number of such effective RCSJ junctions in series. But in the fully ohmic regime at large $I_{\textrm{b}}$, the resistance between the grains would come into play and this could be seen to effectively increase $N$ in Eq.~\ref{eq:icr}, which consequently would decrease $I_c R$ and thereby increase $\Omega$. Thus we believe that the estimate $\Omega \sim 0.005$ calculated from its defining relation should only be a lower bound.

\begin{figure*}
\includegraphics[width=.9\columnwidth]{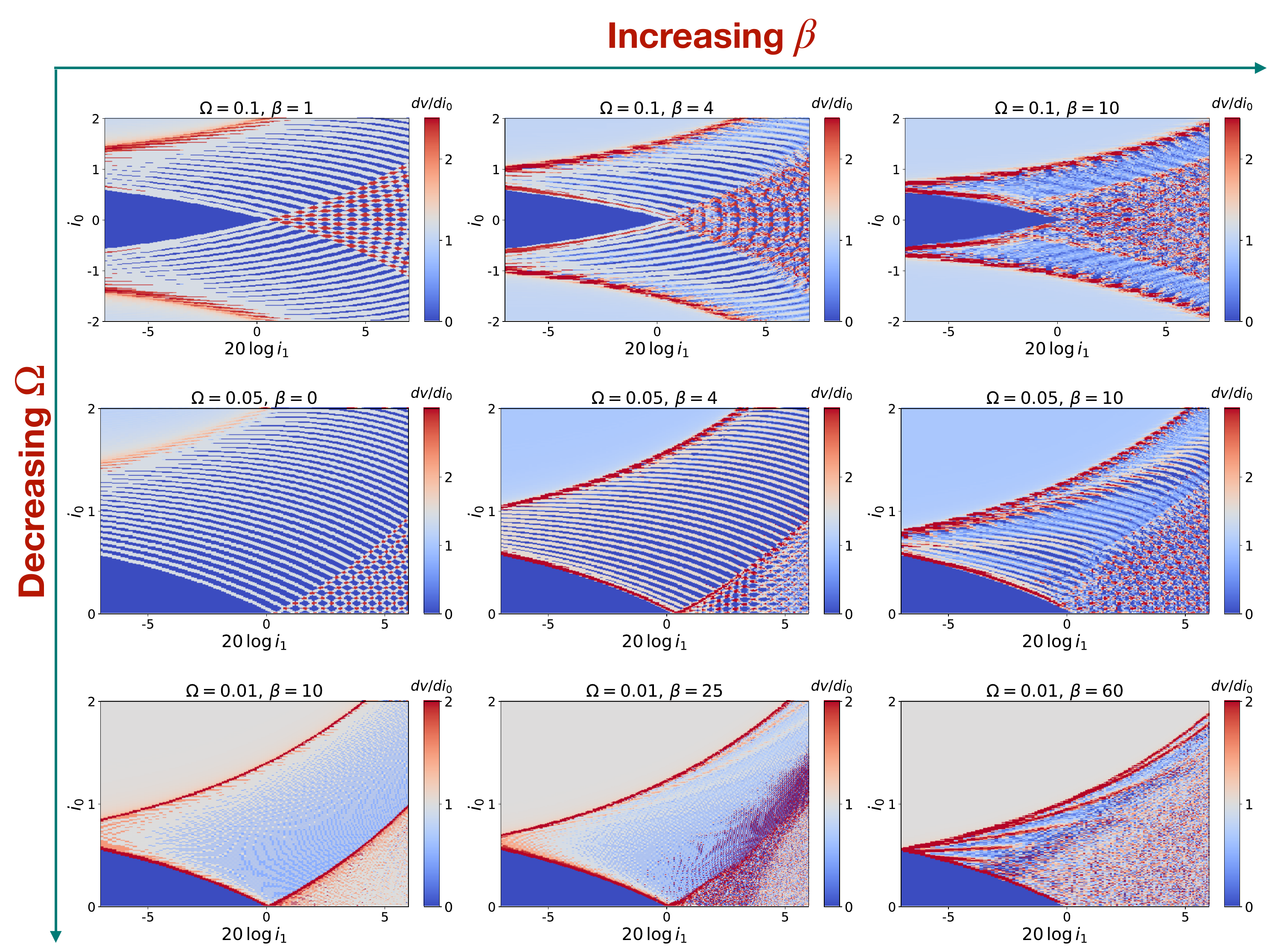}
\caption{\label{fig:rcsj_survey}\justifying Survey of Shapiro maps obtained from the RCSJ model in the $\beta-\Omega$ parameter space. For the cases starting from the  second row, only half of the map is calculated to reduce calculation time.}
\end{figure*}

The direct estimation of the parameter $\beta$ for the RCSJ junctions is complicated since the capacitance $C$ would depend on various factors. We attempt to qualitatively infer $\beta$. If $\beta \ll 1$, then in this overdamped case, the Shapiro steps should extend much above the critical current, unlike in the experiment. On the other hand, if $\beta \gg 1$, then the Shapiro steps would be hardly visible. Both these features could be seen in Fig.~\ref{fig:rcsj_survey}. Based on these considerations we believe that $\beta \sim 1$ in our case.  Next, using the relation $\beta \Omega = 2\pi RC f_{\textrm{RF}}$, and using the inferred $\Omega\sim 0.05$ and $R\sim 2\, \textrm{k}\Omega$ (assuming, $N_c=N_y\times N_z$, we get $I_c\sim 5$ nA and then from $I_cR\sim 0.01$ mV, we get $R$), we get $C\sim 100$ fF. We then see that $E_J/E_c\sim 10$. This large $E_J/E_c$ ratio is consistent with the fact that at low RF powers, the sample is superconducting.

\section{Phenomenological exploration of reentrant superconductivity using the RCSJ model}

In the main text, we attribute the reentrance of superconductivity to a decrease of the charging energy $E_C$ with temperature $T$ via the mechanism discussed in Ref.~\cite{efetov1980phase}. The Efetov model predicts reentrance from an insulating phase to a superconducting phase; here we phenomenologically explore the possibility of reentrance using the RCSJ model, which cannot capture the insulating state. We phenomenologically show that reentrance from a normal phase to a superconducting phase is possible, by assuming that the primary effect of increasing $T$ is to reduce $E_C$ and $E_J$. The reduction of $E_C$ and $E_J$ leads to the reduction of the plasma frequency $\omega_p \sim \sqrt{E_J E_C}$. To study the physics associated with the reduction  of $\omega_p$, it is better to express the RCSJ equation in a dimensionless form where the drive frequency is scaled by the plasma frequency. This is done using the dimensionless time coordinate $\tau=\omega_p t$ in Eq.~\ref{eq:phi_eom} to get,
\begin{equation}
    \partial_\tau^2\phi + \sigma \partial_\tau \phi + \sin \phi =i_0 +i_1 \sin \tilde{\Omega} \tau.
\end{equation}
Here $i_0=I_{\textrm{b}}/I_c$ and $i_{1}=I_{\textrm{RF}}/I_c$ are the scaled currents. The parameter $\sigma$ is the damping parameter and is related to the parameter $\beta$ in the Stewart-McCumber form by $\sigma =1/\sqrt{\beta}$.
The frequency $\tilde{\Omega}$ is scaled by the plasma frequency, i.e. $\tilde{\Omega}=\omega/\omega_p$; $\tilde{\Omega}$ is related to the $\Omega$ in the Stewart-McCumber form by $\tilde{\Omega}=\Omega/\sigma$.

\begin{figure*}
\centering
\includegraphics[width=.9\columnwidth]{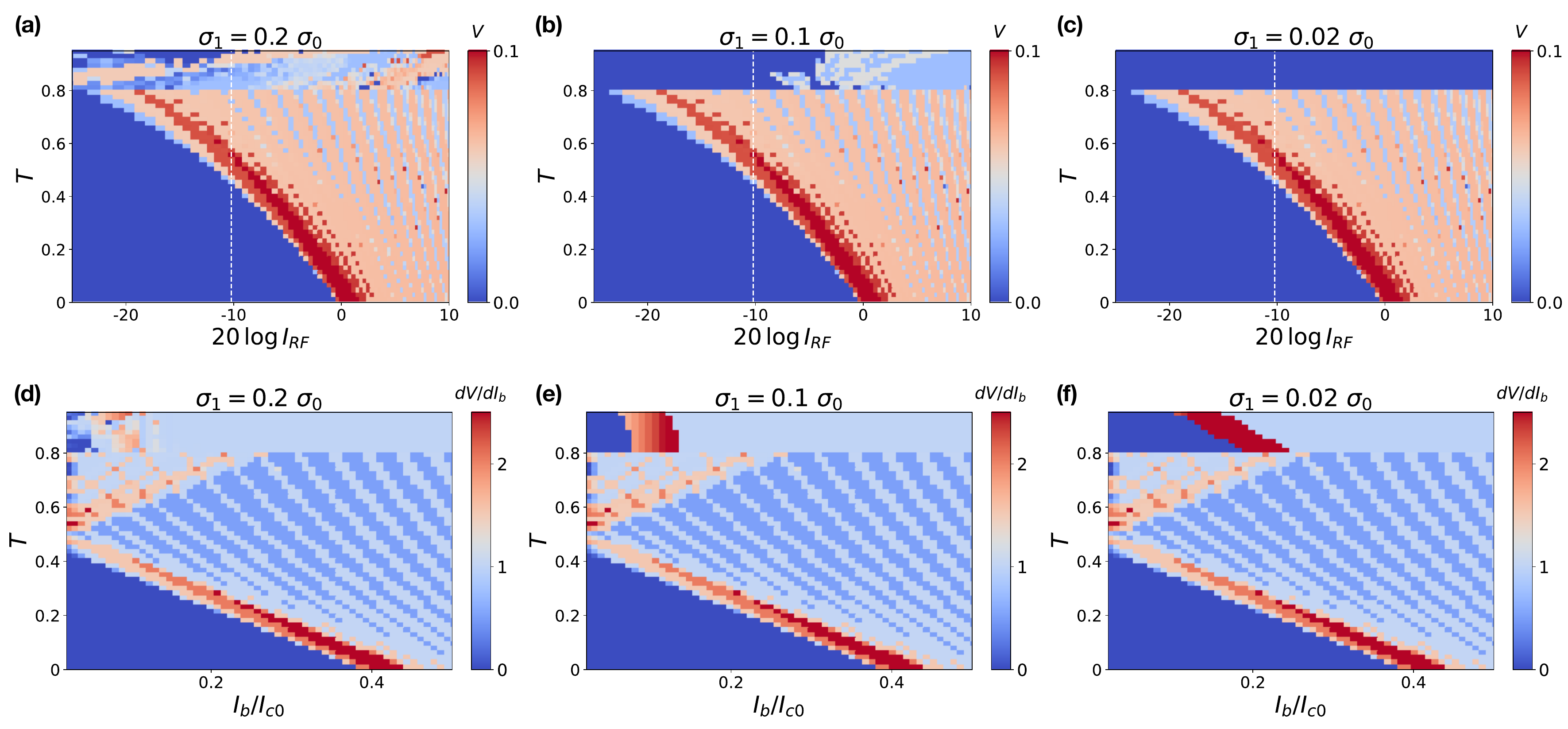}
\caption{\label{fig:rcsj_reentrance} \justifying Exploring reentrance within the RCSJ description, by assuming temperature modified parameters as discussed in the text. The top panels show the case of a small $I_{\mathrm{b}}/I_{c0}=0.05$. The bottom panels show the case with a fixed $I_{\mathrm{RF}}/I_{c0}=0.6$ (indicated by a dashed white line in the top panels). From left to right the magnitude of abrupt $E_C$ reduction at $T^*=0.8T_c$ is increased; this is captured by the ratio of $\sigma_1/\sigma_0$ (see text).}
\end{figure*}

 For simplicity, we assume a linear reduction of $I_c$ with $T$. The exact nature of how $E_C$ changes with $T$ is not known. For the purpose of exploring what magnitude of change in $E_C$ is required for reentrance to occur in the RCSJ description, we consider an abrupt change in $E_C$ after some $T^*<T_c$. This would result in $\sigma=\sigma_0$ at $T<T^*$ going to $\sigma=\sigma_1$ at $T>T^*$. Note that the corresponding change in $\tilde{\Omega}$ follows from its relation to $\Omega$. We choose $\sigma_0=1$ and $\tilde{\Omega}=0.03$ (corresponding to $\beta=1$ and $\Omega=0.03$) at $T=0$, which are close to the parameters estimated in the previous section. We also choose $T^*=0.8 T_c$. The effect of $\sigma_1$ on reentrance is explored in Fig.~\ref{fig:rcsj_reentrance}. 
A crude qualitative resemblance to the experimental results in Fig.~1b and Fig.~2a is seen. However, we reiterate that the feature associated with the strong insulator is not captured, as it is beyond the scope of the RCSJ description.  
From the results, we find that $\sigma_1\lesssim 0.1 \sigma_0$  is required for reentrance. Thus
$E_C$ should decrease by around 3 orders of magnitude for the reentrance to occur within the RCSJ description for the estimated parameter regime of the experiment. 

\section{Extended data and discussions}
\subsection{Reentrant superconductivity}
\begin{figure*}
\begin{subcaptiongroup}
\includegraphics[width=0.95\columnwidth]{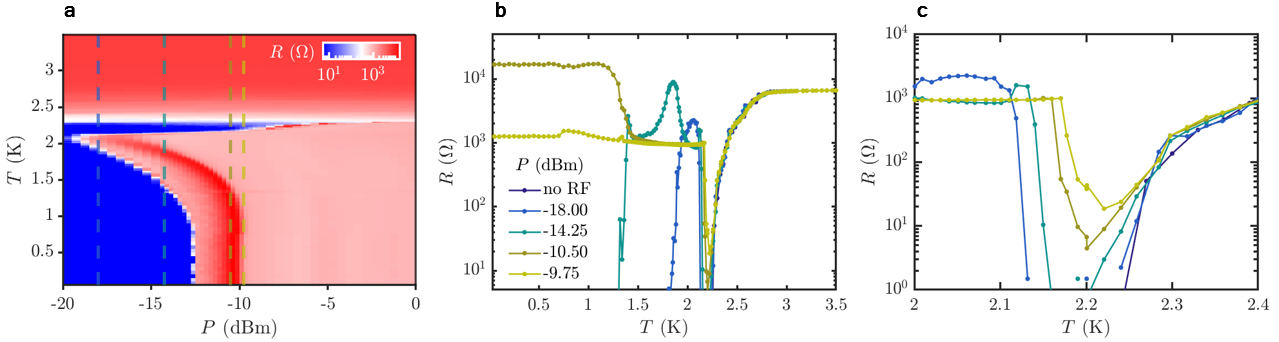}
\phantomcaption\label{fig:RT_linecuts_a}
\phantomcaption\label{fig:RT_linecuts_b}
\phantomcaption\label{fig:RT_linecuts_c}
\end{subcaptiongroup}
\captionsetup{subrefformat=parens}
\caption{\label{fig:RT_linecuts} \justifying
\subref*{fig:RT_linecuts_a} The $R\,(T,P)$ phase diagram presented in Fig. 1b in the main text at frequency $f_{\mathrm{RF}}=24.5$ MHz. \subref*{fig:RT_linecuts_b} $R\,(T)$ linecuts at selected RF power levels indicated by dashed lines in (a). \subref*{fig:RT_linecuts_c} The $R\,(T)$ linecuts in the reentrance temperature region SC 2.}
\end{figure*}

To examine the reentrant superconductivity shown in the phase diagram of Fig. 1b in the main text, we analyze five representative line cuts taken at constant power levels, indicated by dashed lines in Fig. \ref{fig:RT_linecuts_a}. The corresponding resistance versus temperature ($R$ vs. $T$) curves for each power level are presented in Fig. \ref{fig:RT_linecuts_b}. A consistent color scheme is used across both figures to denote the different power levels, as referenced to in Fig. \ref{fig:RT_linecuts_a}. For comparison, we also include the $R$ vs. $T$ trace measured without applied RF power, which exhibits a single superconducting transition at a critical temperature of $T_c = 2.25$ K. 

Figure \ref{fig:RT_linecuts_c} provides a closer view of the $R(T)$ curves in the region where reentrant superconductivity emerges. At $P = -18$ dBm, three distinct transitions are observed: the resistance drops to zero near $T \approx T_c$, rises again at $T = 2.13$ K, and then falls back to zero at $T = 1.8$ K. At a higher power level ($P = -10.5$ dBm), two superconducting transitions are seen, indicating a narrowing of the reentrant region with increasing RF power.

\subsection{The insulating state}
\subsubsection{Estimating the normal state resistance}
To estimate the normal-state resistance, we rely on high-current measurements rather than resistance data taken above the superconducting transition temperature ($T_c$). This is because measurements above $T_c$ include significant contributions from the contact regions, due to the quasi-four-probe configuration of our setup. Instead, we take advantage of the device geometry, which features relatively wide contact regions.
As described in Ref. \cite{PhysRevApplied.4.024021}, upon increasing the DC current $I_\textrm{b}$ beyond $I_c$, defined in Fig. \ref{fig:insulator_a}, the nanobridge and parts of the contacts switch to the resistive state. When $I_\textrm{b}$ is reduced towards $I_r$ (also defined in Fig. \ref{fig:insulator_a}) the contacts return to the superconducting state first, followed by the nanobridge at $I_\textrm{b}=I_{r}$. Thus we estimate $R_\mathrm{N}$ as $dV/dI_{\textrm{b}}$ obtained right before $I_\textrm{b}=I_{r}$, where we assume the nanobridge is in the normal state while the contacts are superconducting.  While it is possible that parts of the contact pads also remain resistive under these conditions, potentially increasing the measured resistance, we treat this value as an upper bound for the true normal-state resistance. Using this procedure we obtain an upper limit to the normal-state resistance, $R_\mathrm{N}=1.55\ \mathrm{k\Omega}$. 
\subsubsection{Study the temperature dependence of the insulating state}
In Fig. \ref{fig:insulator_b} we demonstrate that in the insulating (I) state, the resistance follows a sharp exponential rise beyond ten times $R_\mathrm{N}$, then saturates at low temperatures. The observed exponential rise is stronger than that observed for electron localization triggered by Coulomb interactions in insulating granular aluminum films \cite{PhysRevLett.44.1150}. At low temperatures the resistance saturates close to $50\ \mathrm{k\Omega}$. We believe that this saturation is not intrinsic and it is due to the experimental limitations in our measurement system. We demonstrate that in the next paragraph 
\subsubsection{Origin of the low temperature saturation}
\begin{figure*}
\begin{subcaptiongroup}
\begin{center}
\includegraphics[width=1\columnwidth]
{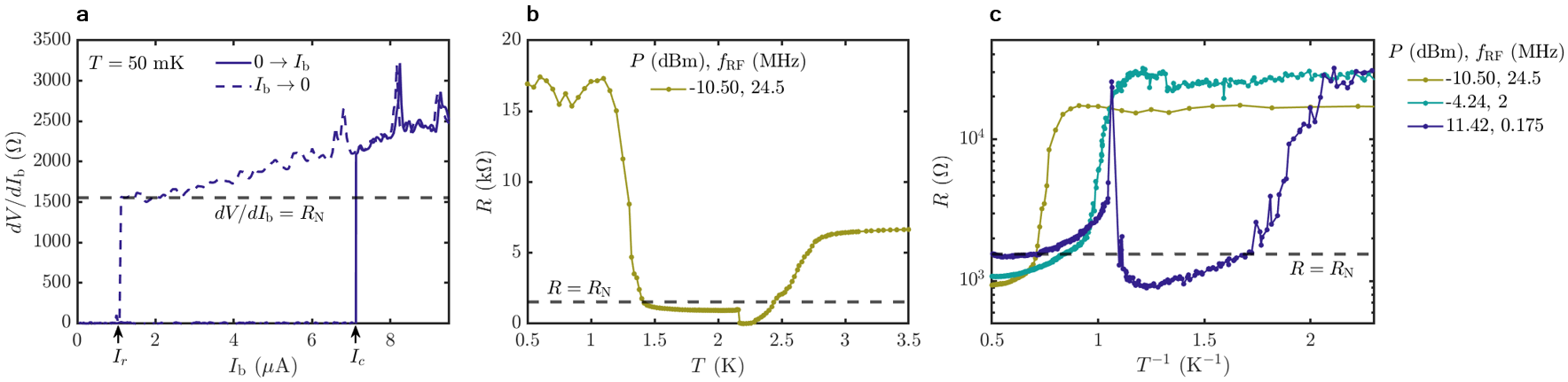}
\end{center}
\phantomcaption\label{fig:insulator_a}
\phantomcaption\label{fig:insulator_b}
\phantomcaption\label{fig:insulator_c}
\end{subcaptiongroup}
\captionsetup{subrefformat=parens}
\caption{\label{fig:insulator} \justifying
\subref*{fig:insulator_a} $dV/dI_\mathrm{b}\,(I_\mathrm{b})$ measured at opposite current sweep directions in the absence of RF power. \subref*{fig:insulator_b} $R\,(T)$ measured in the presence of RF power of $P_{\mathrm{I}}=-10.5$ dBm and $f_{\mathrm{RF}}=24.5$ MHz, showing the insulating state in comparison to the normal state resistance $R_\mathrm{N}$. \subref*{fig:insulator_c} Exploration of the RF frequency dependence of the insulating state. $R\,(T)$ curves are shown for selected RF frequencies and their corresponding input RF powers.}
\end{figure*}

Figure \ref{fig:insulator_c} examines the frequency dependence of resistance in the insulating state. For each RF frequency ($f_{\mathrm{RF}}$), we measure resistance ($R$) as a function of input power ($P$) at base temperature and identify the power level, $P_{\mathrm{I}}(f_{\mathrm{RF}})$, at which $R$ reaches its maximum. While the actual RF power delivered to the sample may vary with frequency, selecting the peak resistance point allows us to assume that the effective RF power within the device is approximately frequency-independent.

At lower RF frequencies, we observe a lower saturation temperature. One possible explanation is that as the device enters the insulating regime and resistance increases, more power is dissipated, potentially heating the electrons and limiting how cold the electron temperature can get. Another contributing factor may be limitations in our measurement setup, particularly the RF bias tee filters, which include 50 k$\Omega$ resistors—comparable to the resistance values observed at saturation. These constraints suggest that, in the absence of such limitations, the resistance could have continued to increase with further cooling.  

\subsection{Finite voltage width of the giant Shapiro steps}
\begin{figure*}[!h]
\begin{subcaptiongroup}
\includegraphics[width=0.75\columnwidth]{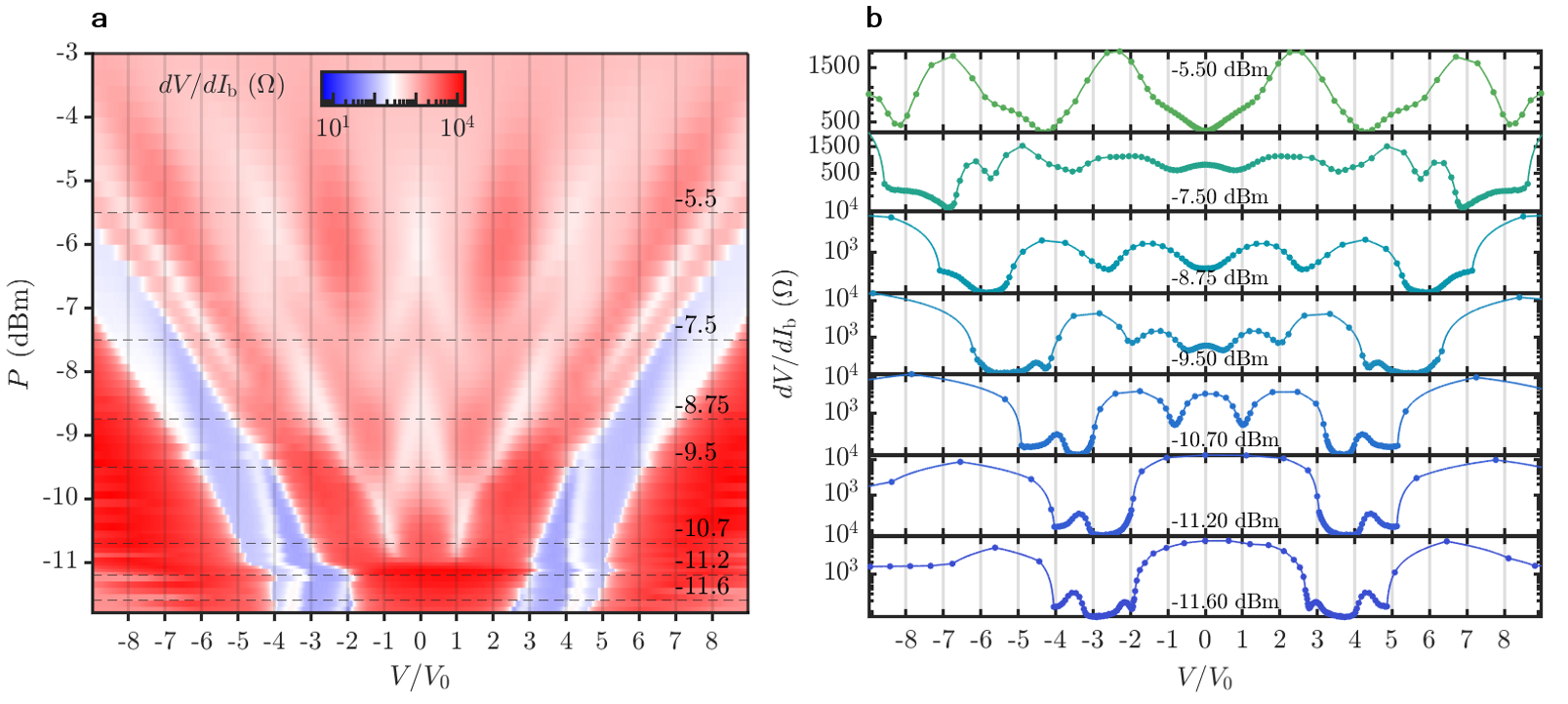}
\phantomcaption\label{fig:giantsteps_a}
\phantomcaption\label{fig:giantsteps_b}
\end{subcaptiongroup}
\captionsetup{subrefformat=parens}
\caption{\label{fig:giantsteps} \justifying \subref*{fig:giantsteps_a} $dV/dI_{\mathrm{b}}$ as a function of $V/V_0$ and $P$ at $f_{\mathrm{RF}}=67.5\ \mathrm{MHz}$, where $V_0=60\ \mathrm{\mu V}$. \subref*{fig:giantsteps_b} $dV/dI_{\mathrm{b}}$ vs. $V/V_0$ curves obtained at constant power levels indicated by dashed lines in \subref*{fig:giantsteps_a}.}
\end{figure*}

Here we demonstrate that the observed giant Shapiro steps are characterized by a finite voltage width. In Fig. \ref{fig:giantsteps_a} we show $dV/dI_\mathrm{b}$ as a function of $P$ and the normalized voltage $V/V_0$, at $f_{\mathrm{RF}}=67.5$ MHz. The voltage $V$ is obtained by integrating the measured $dV/dI_\mathrm{b}$ with respect to $I_\mathrm{b}$. $V_0=60\ \mathrm{\mu V}$ is the typical quantization voltage obtained in Fig. 3b in the main text. As can be seen, the quantized voltage is significantly smeared. The voltage steps are reflected in bright regions in the $dV/dI_\mathrm{b}$ color map. In the case of a uniform JJ array, these bright regions should follow straight vertical lines at integer values of $V/V_0$ (indicated by solid lines in Fig. \ref{fig:giantsteps_a} for comparison). In contrast, here the bright regions form a diagonal pattern, smeared towards non-integer $V/V_0$ values. This finite voltage width is further demonstrated in Fig. \ref{fig:giantsteps_b}, where we analyze several representative line cuts taken at constant power levels, indicated by dashed lines in Fig. \ref{fig:giantsteps_a}. At $P=-10.7$ dBm, we see a minimum of $dV/dI_\mathrm{b}$ at $V/V_0\simeq1$, while for $P=-9.5$ dBm this minimum appears at $V/V_0\simeq0.5$ due to the smearing. This smearing may result from distribution of junction properties in our naturally occurring JJ array. This disorder, however, may not be sufficient to completely diminish the giant steps observed; therefore, we expect it to be minimal.

\subsection{Absence of RF overheating}
The presence of RF power may create dissipation and electron overheating \cite{DeCecco2016}. It appears as a discontinuous jump of $I_{c,\mathrm{eff}}$ with increasing $I_{\mathrm{RF}}$. However, our naturally occurring array is controlled to exhibit negligible overheating as it is tunable at low $f_{\mathrm{RF}}$, since the dissipated power is proportional to $f_{\mathrm{RF}}^2$. This is evident by the continuous suppression of $I_{c,\mathrm{eff}}$ with increasing $P$ (Fig. 1a) and the absence of hysteresis of $R$ over $P$ sweeps for frequencies $f_{\mathrm{RF}}<25\ \mathrm{MHz}$ (not shown). This contrasts with operation at high frequencies, where the discontinuity is visible, as can be seen in Fig 3c in the main text for $f_\mathrm{RF}=67.5$ MHz. Therefore, we suggest that the observed superconductor-to-insulator transition and reentrant superconductivity can be interpreted in the regime of negligible overheating.

\begin{figure*}[!h]
\includegraphics[width=0.5\columnwidth]{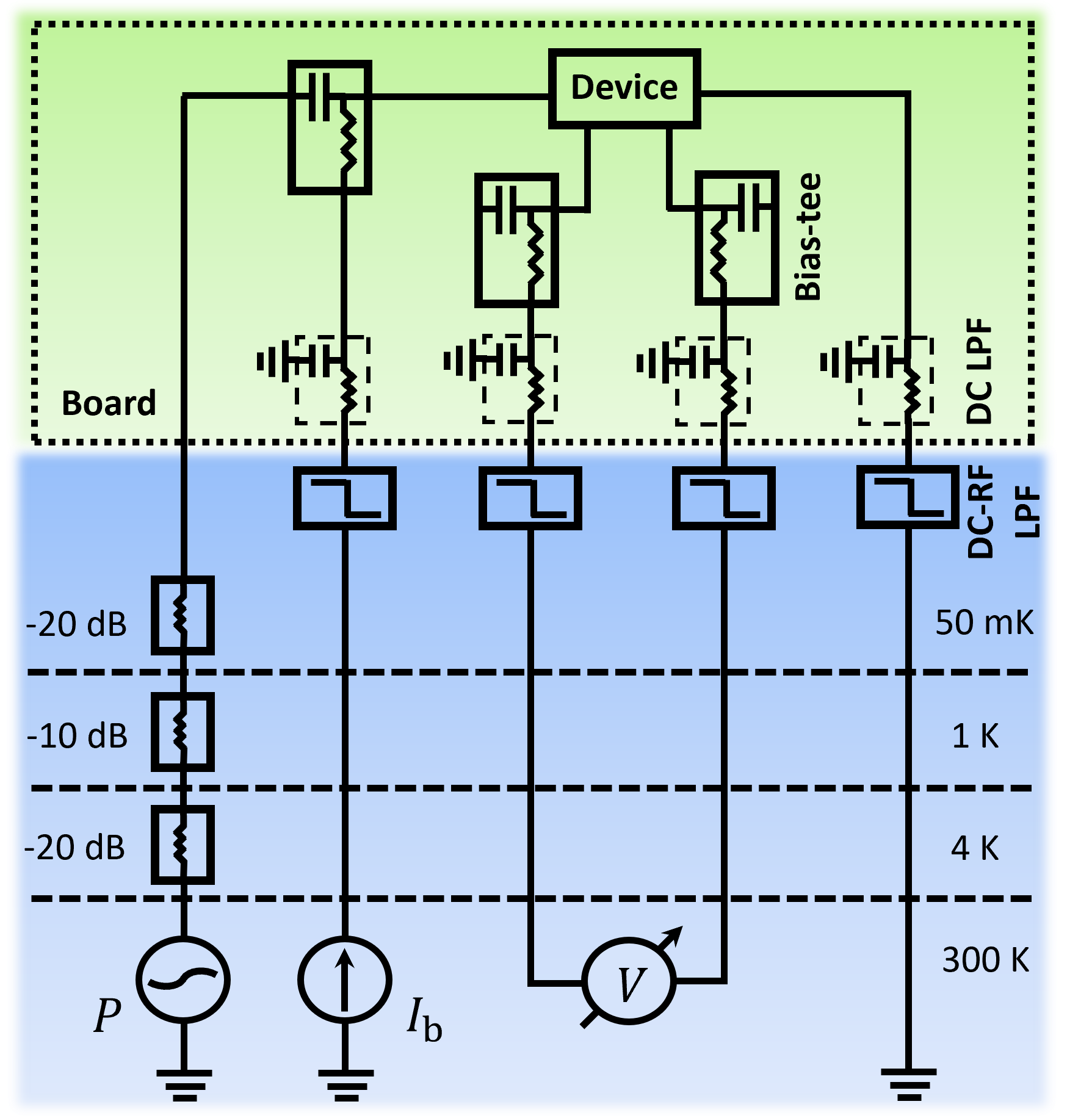}
\caption{\label{fig:setup} The measurement setup used in the experiment.}
\end{figure*}

\bibliography{SuppRef}

\begin{thebibliography}{11}%
\makeatletter
\providecommand \@ifxundefined [1]{%
 \@ifx{#1\undefined}
}%
\providecommand \@ifnum [1]{%
 \ifnum #1\expandafter \@firstoftwo
 \else \expandafter \@secondoftwo
 \fi
}%
\providecommand \@ifx [1]{%
 \ifx #1\expandafter \@firstoftwo
 \else \expandafter \@secondoftwo
 \fi
}%
\providecommand \natexlab [1]{#1}%
\providecommand \enquote  [1]{``#1''}%
\providecommand \bibnamefont  [1]{#1}%
\providecommand \bibfnamefont [1]{#1}%
\providecommand \citenamefont [1]{#1}%
\providecommand \href@noop [0]{\@secondoftwo}%
\providecommand \href [0]{\begingroup \@sanitize@url \@href}%
\providecommand \@href[1]{\@@startlink{#1}\@@href}%
\providecommand \@@href[1]{\endgroup#1\@@endlink}%
\providecommand \@sanitize@url [0]{\catcode `\\12\catcode `\$12\catcode `\&12\catcode `\#12\catcode `\^12\catcode `\_12\catcode `\%12\relax}%
\providecommand \@@startlink[1]{}%
\providecommand \@@endlink[0]{}%
\providecommand \url  [0]{\begingroup\@sanitize@url \@url }%
\providecommand \@url [1]{\endgroup\@href {#1}{\urlprefix }}%
\providecommand \urlprefix  [0]{URL }%
\providecommand \Eprint [0]{\href }%
\providecommand \doibase [0]{https://doi.org/}%
\providecommand \selectlanguage [0]{\@gobble}%
\providecommand \bibinfo  [0]{\@secondoftwo}%
\providecommand \bibfield  [0]{\@secondoftwo}%
\providecommand \translation [1]{[#1]}%
\providecommand \BibitemOpen [0]{}%
\providecommand \bibitemStop [0]{}%
\providecommand \bibitemNoStop [0]{.\EOS\space}%
\providecommand \EOS [0]{\spacefactor3000\relax}%
\providecommand \BibitemShut  [1]{\csname bibitem#1\endcsname}%
\let\auto@bib@innerbib\@empty
\bibitem [{\citenamefont {Koch}\ \emph {et~al.}(2007)\citenamefont {Koch}, \citenamefont {Yu}, \citenamefont {Gambetta}, \citenamefont {Houck}, \citenamefont {Schuster}, \citenamefont {Majer}, \citenamefont {Blais}, \citenamefont {Devoret}, \citenamefont {Girvin},\ and\ \citenamefont {Schoelkopf}}]{koch2007charge}%
  \BibitemOpen
  \bibfield  {author} {\bibinfo {author} {\bibfnamefont {J.}~\bibnamefont {Koch}}, \bibinfo {author} {\bibfnamefont {T.~M.}\ \bibnamefont {Yu}}, \bibinfo {author} {\bibfnamefont {J.}~\bibnamefont {Gambetta}}, \bibinfo {author} {\bibfnamefont {A.~A.}\ \bibnamefont {Houck}}, \bibinfo {author} {\bibfnamefont {D.~I.}\ \bibnamefont {Schuster}}, \bibinfo {author} {\bibfnamefont {J.}~\bibnamefont {Majer}}, \bibinfo {author} {\bibfnamefont {A.}~\bibnamefont {Blais}}, \bibinfo {author} {\bibfnamefont {M.~H.}\ \bibnamefont {Devoret}}, \bibinfo {author} {\bibfnamefont {S.~M.}\ \bibnamefont {Girvin}},\ and\ \bibinfo {author} {\bibfnamefont {R.~J.}\ \bibnamefont {Schoelkopf}},\ }\bibfield  {title} {\bibinfo {title} {Charge-insensitive qubit design derived from the cooper pair box},\ }\href {https://doi.org/10.1103/PhysRevA.76.042319} {\bibfield  {journal} {\bibinfo  {journal} {Phys. Rev. A}\ }\textbf {\bibinfo {volume} {76}},\ \bibinfo {pages} {042319} (\bibinfo {year} {2007})}\BibitemShut {NoStop}%
\bibitem [{\citenamefont {Likharev}\ and\ \citenamefont {Zorin}(1985)}]{likharev1985bloch}%
  \BibitemOpen
  \bibfield  {author} {\bibinfo {author} {\bibfnamefont {K.~K.}\ \bibnamefont {Likharev}}\ and\ \bibinfo {author} {\bibfnamefont {A.~B.}\ \bibnamefont {Zorin}},\ }\bibfield  {title} {\bibinfo {title} {Theory of the bloch-wave oscillations in small josephson junctions},\ }\href {https://doi.org/10.1007/BF00683782} {\bibfield  {journal} {\bibinfo  {journal} {Journal of Low Temperature Physics}\ }\textbf {\bibinfo {volume} {59}},\ \bibinfo {pages} {347} (\bibinfo {year} {1985})}\BibitemShut {NoStop}%
\bibitem [{\citenamefont {Penttil\"a}\ \emph {et~al.}(1999)\citenamefont {Penttil\"a}, \citenamefont {Parts}, \citenamefont {Hakonen}, \citenamefont {Paalanen},\ and\ \citenamefont {Sonin}}]{penttila1999superconductor}%
  \BibitemOpen
  \bibfield  {author} {\bibinfo {author} {\bibfnamefont {J.~S.}\ \bibnamefont {Penttil\"a}}, \bibinfo {author} {\bibfnamefont {U.}~\bibnamefont {Parts}}, \bibinfo {author} {\bibfnamefont {P.~J.}\ \bibnamefont {Hakonen}}, \bibinfo {author} {\bibfnamefont {M.~A.}\ \bibnamefont {Paalanen}},\ and\ \bibinfo {author} {\bibfnamefont {E.~B.}\ \bibnamefont {Sonin}},\ }\bibfield  {title} {\bibinfo {title} {``superconductor-insulator transition'' in a single josephson junction},\ }\href {https://doi.org/10.1103/PhysRevLett.82.1004} {\bibfield  {journal} {\bibinfo  {journal} {Phys. Rev. Lett.}\ }\textbf {\bibinfo {volume} {82}},\ \bibinfo {pages} {1004} (\bibinfo {year} {1999})}\BibitemShut {NoStop}%
\bibitem [{\citenamefont {Takahide}\ \emph {et~al.}(2000)\citenamefont {Takahide}, \citenamefont {Yagi}, \citenamefont {Kanda}, \citenamefont {Ootuka},\ and\ \citenamefont {Kobayashi}}]{takahide2000superconductor}%
  \BibitemOpen
  \bibfield  {author} {\bibinfo {author} {\bibfnamefont {Y.}~\bibnamefont {Takahide}}, \bibinfo {author} {\bibfnamefont {R.}~\bibnamefont {Yagi}}, \bibinfo {author} {\bibfnamefont {A.}~\bibnamefont {Kanda}}, \bibinfo {author} {\bibfnamefont {Y.}~\bibnamefont {Ootuka}},\ and\ \bibinfo {author} {\bibfnamefont {S.-i.}\ \bibnamefont {Kobayashi}},\ }\bibfield  {title} {\bibinfo {title} {Superconductor-insulator transition in a two-dimensional array of resistively shunted small josephson junctions},\ }\href {https://doi.org/10.1103/PhysRevLett.85.1974} {\bibfield  {journal} {\bibinfo  {journal} {Phys. Rev. Lett.}\ }\textbf {\bibinfo {volume} {85}},\ \bibinfo {pages} {1974} (\bibinfo {year} {2000})}\BibitemShut {NoStop}%
\bibitem [{\citenamefont {G\'omez-Le\'on}\ and\ \citenamefont {Platero}(2013)}]{gomez2013floquet}%
  \BibitemOpen
  \bibfield  {author} {\bibinfo {author} {\bibfnamefont {A.}~\bibnamefont {G\'omez-Le\'on}}\ and\ \bibinfo {author} {\bibfnamefont {G.}~\bibnamefont {Platero}},\ }\bibfield  {title} {\bibinfo {title} {Floquet-bloch theory and topology in periodically driven lattices},\ }\href {https://doi.org/10.1103/PhysRevLett.110.200403} {\bibfield  {journal} {\bibinfo  {journal} {Phys. Rev. Lett.}\ }\textbf {\bibinfo {volume} {110}},\ \bibinfo {pages} {200403} (\bibinfo {year} {2013})}\BibitemShut {NoStop}%
\bibitem [{\citenamefont {Grossmann}\ \emph {et~al.}(1991)\citenamefont {Grossmann}, \citenamefont {Dittrich}, \citenamefont {Jung},\ and\ \citenamefont {H\"anggi}}]{grossmann1991coherent}%
  \BibitemOpen
  \bibfield  {author} {\bibinfo {author} {\bibfnamefont {F.}~\bibnamefont {Grossmann}}, \bibinfo {author} {\bibfnamefont {T.}~\bibnamefont {Dittrich}}, \bibinfo {author} {\bibfnamefont {P.}~\bibnamefont {Jung}},\ and\ \bibinfo {author} {\bibfnamefont {P.}~\bibnamefont {H\"anggi}},\ }\bibfield  {title} {\bibinfo {title} {Coherent destruction of tunneling},\ }\href {https://doi.org/10.1103/PhysRevLett.67.516} {\bibfield  {journal} {\bibinfo  {journal} {Phys. Rev. Lett.}\ }\textbf {\bibinfo {volume} {67}},\ \bibinfo {pages} {516} (\bibinfo {year} {1991})}\BibitemShut {NoStop}%
\bibitem [{\citenamefont {Ravindran}\ \emph {et~al.}(1996)\citenamefont {Ravindran}, \citenamefont {Gómez}, \citenamefont {Li}, \citenamefont {Herbert}, \citenamefont {Lukens}, \citenamefont {Jun}, \citenamefont {Elhamri}, \citenamefont {Newrock},\ and\ \citenamefont {Mast}}]{Ravindran1996}%
  \BibitemOpen
  \bibfield  {author} {\bibinfo {author} {\bibfnamefont {K.}~\bibnamefont {Ravindran}}, \bibinfo {author} {\bibfnamefont {L.}~\bibnamefont {Gómez}}, \bibinfo {author} {\bibfnamefont {R.}~\bibnamefont {Li}}, \bibinfo {author} {\bibfnamefont {S.}~\bibnamefont {Herbert}}, \bibinfo {author} {\bibfnamefont {P.}~\bibnamefont {Lukens}}, \bibinfo {author} {\bibfnamefont {Y.}~\bibnamefont {Jun}}, \bibinfo {author} {\bibfnamefont {S.}~\bibnamefont {Elhamri}}, \bibinfo {author} {\bibfnamefont {R.}~\bibnamefont {Newrock}},\ and\ \bibinfo {author} {\bibfnamefont {D.}~\bibnamefont {Mast}},\ }\bibfield  {title} {\bibinfo {title} {Frequency dependence of giant shapiro steps in ordered and site-disordered proximity-coupled josephson-junction arrays},\ }\href {https://doi.org/10.1103/PhysRevB.53.5141} {\bibfield  {journal} {\bibinfo  {journal} {Phys. Rev. B}\ }\textbf {\bibinfo {volume} {53}},\ \bibinfo {pages} {5141} (\bibinfo {year} {1996})}\BibitemShut {NoStop}%
\bibitem [{\citenamefont {Efetov}(1980)}]{efetov1980phase}%
  \BibitemOpen
  \bibfield  {author} {\bibinfo {author} {\bibfnamefont {K.}~\bibnamefont {Efetov}},\ }\bibfield  {title} {\bibinfo {title} {Phase transition in granulated superconductors},\ }\href@noop {} {\bibfield  {journal} {\bibinfo  {journal} {Sov. Phys.-JETP (Engl. Transl.);(United States)}\ }\textbf {\bibinfo {volume} {51}} (\bibinfo {year} {1980})}\BibitemShut {NoStop}%
\bibitem [{\citenamefont {Hazra}\ \emph {et~al.}(2015)\citenamefont {Hazra}, \citenamefont {Kirtley},\ and\ \citenamefont {Hasselbach}}]{PhysRevApplied.4.024021}%
  \BibitemOpen
  \bibfield  {author} {\bibinfo {author} {\bibfnamefont {D.}~\bibnamefont {Hazra}}, \bibinfo {author} {\bibfnamefont {J.~R.}\ \bibnamefont {Kirtley}},\ and\ \bibinfo {author} {\bibfnamefont {K.}~\bibnamefont {Hasselbach}},\ }\bibfield  {title} {\bibinfo {title} {Retrapping current in bridge-type nano-squids},\ }\href {https://doi.org/10.1103/PhysRevApplied.4.024021} {\bibfield  {journal} {\bibinfo  {journal} {Phys. Rev. Appl.}\ }\textbf {\bibinfo {volume} {4}},\ \bibinfo {pages} {024021} (\bibinfo {year} {2015})}\BibitemShut {NoStop}%
\bibitem [{\citenamefont {Deutscher}\ \emph {et~al.}(1980)\citenamefont {Deutscher}, \citenamefont {Bandyopadhyay}, \citenamefont {Chui}, \citenamefont {Lindenfeld}, \citenamefont {McLean},\ and\ \citenamefont {Worthington}}]{PhysRevLett.44.1150}%
  \BibitemOpen
  \bibfield  {author} {\bibinfo {author} {\bibfnamefont {G.}~\bibnamefont {Deutscher}}, \bibinfo {author} {\bibfnamefont {B.}~\bibnamefont {Bandyopadhyay}}, \bibinfo {author} {\bibfnamefont {T.}~\bibnamefont {Chui}}, \bibinfo {author} {\bibfnamefont {P.}~\bibnamefont {Lindenfeld}}, \bibinfo {author} {\bibfnamefont {W.~L.}\ \bibnamefont {McLean}},\ and\ \bibinfo {author} {\bibfnamefont {T.}~\bibnamefont {Worthington}},\ }\bibfield  {title} {\bibinfo {title} {Transition to localization in granular aluminum films},\ }\href {https://doi.org/10.1103/PhysRevLett.44.1150} {\bibfield  {journal} {\bibinfo  {journal} {Phys. Rev. Lett.}\ }\textbf {\bibinfo {volume} {44}},\ \bibinfo {pages} {1150} (\bibinfo {year} {1980})}\BibitemShut {NoStop}%
\bibitem [{\citenamefont {Cecco}\ \emph {et~al.}(2016)\citenamefont {Cecco}, \citenamefont {Calvez}, \citenamefont {Sacépé}, \citenamefont {Winkelmann},\ and\ \citenamefont {Courtois}}]{DeCecco2016}%
  \BibitemOpen
  \bibfield  {author} {\bibinfo {author} {\bibfnamefont {A.~D.}\ \bibnamefont {Cecco}}, \bibinfo {author} {\bibfnamefont {K.~L.}\ \bibnamefont {Calvez}}, \bibinfo {author} {\bibfnamefont {B.}~\bibnamefont {Sacépé}}, \bibinfo {author} {\bibfnamefont {C.~B.}\ \bibnamefont {Winkelmann}},\ and\ \bibinfo {author} {\bibfnamefont {H.}~\bibnamefont {Courtois}},\ }\bibfield  {title} {\bibinfo {title} {Interplay between electron overheating and ac josephson effect},\ }\href {https://doi.org/10.1103/PHYSREVB.93.180505/ALESSANDROPAPERSNS-16SUPP.PDF} {\bibfield  {journal} {\bibinfo  {journal} {Phys. Rev. B}\ }\textbf {\bibinfo {volume} {93}},\ \bibinfo {pages} {180505} (\bibinfo {year} {2016})}\BibitemShut {NoStop}%
\end{thebibliography}%



\end{document}